\documentclass[aps]{revtex4}

\usepackage{graphics}

\begin{document}
\def\be{\begin{equation}}
\def\ee{\end{equation}}
\def\beqa{\begin{eqnarray}}
\def\eeqa{\end{eqnarray}}
\def\om{\Omega}
\def\vxi{\vec{\xi}}
\def\x{\vec{x}}
\def\xp{\vec{x}^{\, \prime}}
\def\k{\vec{k}}
\def\q{\vec{q}}
\def\e{{\rm e}}
\def\im{{\rm i}}
\def\dd{{\rm d}}
\def\U{{\cal U}}
\def\V{{\cal V}}

\title{ Multi-photon, multi-mode polarization entanglement in parametric down-conversion}
\author{A.~Gatti$^{1}$, R.Zambrini$^{2}$, M.~San Miguel$^{2}$ and
L.~A.~Lugiato$^{1}$}
\address{$^1$ INFM, Dipartimento di Scienze CC FF MM, Universit\'a
dell'Insubria, Via Valleggio 11, 22100 Como, Italy.}
\address{$^2$ 
IMEDEA, Campus Universitat Illes Balears, E-07071
Palma de Mallorca, Spain.}
\begin{abstract}
We study the quantum properties of the polarization of the light produced in type II
spontaneous parametric down-conversion in the framework of a multi-mode model valid in any gain regime.
We show that the the microscopic polarization entanglement of photon pairs survives 
in the high gain regime (multi-photon regime), 
in the form of 
nonclassical correlation of {\em all} the Stokes operators describing polarization degrees of freedom.
\end{abstract}
\maketitle
\section{Introduction}
The quantum properties of light polarization have been widely studied 
in the regime of single photon counts. In comparison, only recently there has been a rise of interest
towards  the quantum properties
of the polarization of macroscopic light beams \cite{Bowen, Korolkova,Bowen2,Bowen3}, mainly due to their potential applications to
the field of quantum information with continuous variables and to
the possibility of mapping the quantum state from light to
atomic media \cite{Polzik}.\\ 
A well-known source of polarization entangled photons is
parametric down-conversion in a type II crystal. Here, a pump field at high frequency is partially converted
into two fields at lower frequency, distinguished by their polarizations. Due to spatial walk-off in the crystal,
the two emission cones are slightly displaced one with respect to the other, and
the the far-field intensity
distribution has the shape of two rings, whose centers are displaced along the walk-off direction,
as e.g. shown by Fig.1. 
\begin{figure}[b]
\centerline{\scalebox{.60}{\includegraphics*{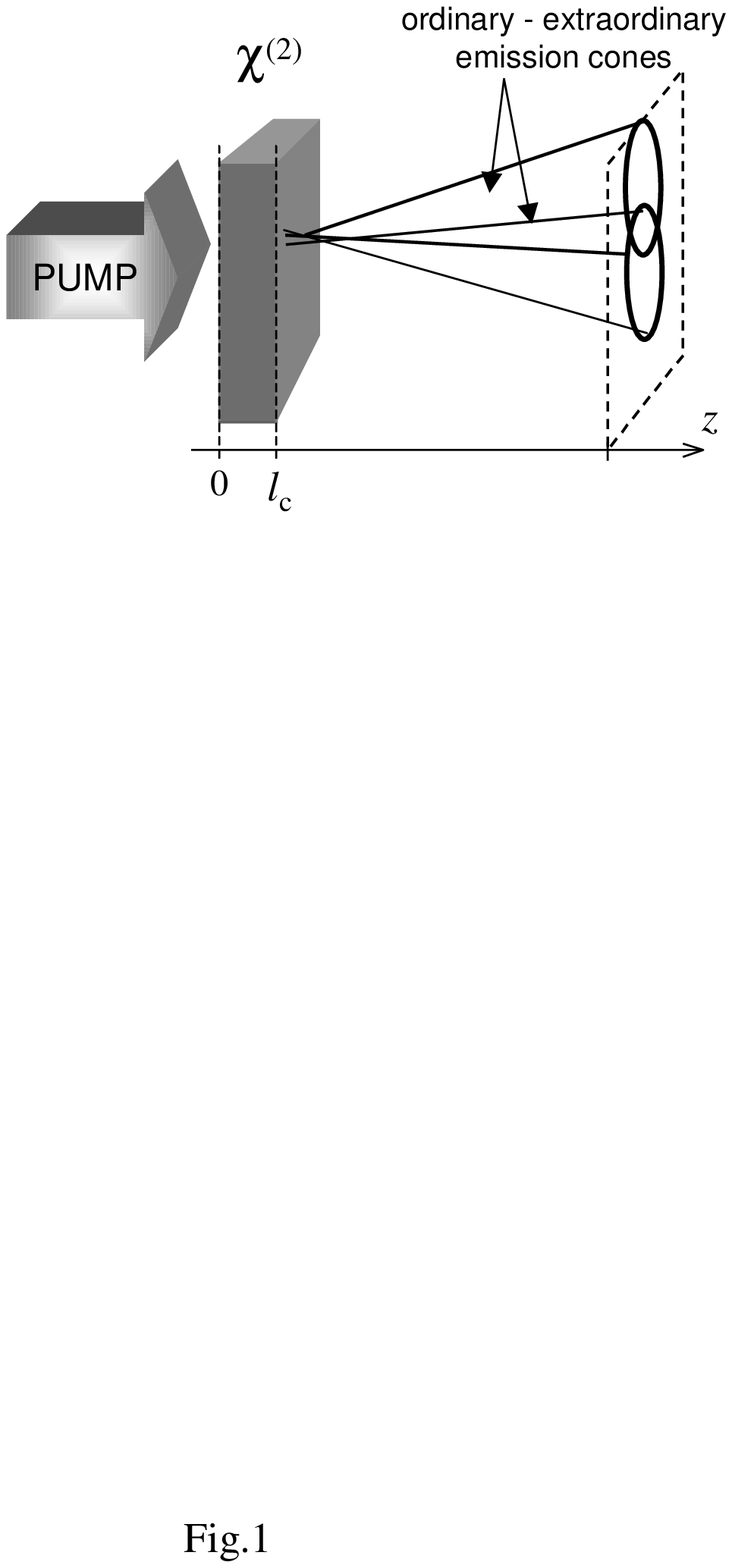}}}
\caption{Parametric down-conversion from a type II crystal showing the two down-conversion cones at degeneracy.}
\label{fig1}
\end{figure}
The two regions where the far-field rings intersect have a very
special role. In the regime of  single photon pair detection, 
the polarization of a photon detected in one of this region is 
completely undetermined. However, once the polarization of one photon has been measured, 
the polarization of the other photon, which propagates at the symmetric position, 
is exactly determined. In other words, when considering photodetection from these regions,  
the two-photon state can be described as the ideal polarization-entangled state\cite{kwiat}. 
Photons produced by this process has become an essential ingredient in many implementations
of quantum imformation schemes (see e.g. \cite{teleportation,cryptography}). 
 
The question that we address in this paper is whether the microscopic 
photon polarization entaglement leaves any trace in the regime
of high parametric down-conversion efficiency, where the number of down-converted photons
can be rather large \cite{Ottavia}, and in which form.

To this end parametric down-conversion is described in the framework of a multi-mode model,
valid for any gain regime, which includes typical effects present in a realistic crystal, as
diffraction and spatio-temporal walk-off.
Quantum-optical polarization properties of the down-converted light  are
described within the formalism of Stokes operators. 
These operators obey to angular
momentum-like commutation rules, and the associated observables are in general non compatible.
We define a local version of Stokes operators and study the quantum correlation
between  Stokes operators measured from symmetric portions of the beam cross-section in the far field zone. 
In the regions where the two down-conversion cones intersect we find that all the Stokes operators
are correlated at the quantum level. Although the
light is completely unpolarized and Stokes operators are very noisy, a measurement of a Stokes
parameter in one of these regions in any polarization basis determines the value of the Stokes parameter
in the symmetric region within an uncertainty much below the standard quantum limit.

A continuous variable polarization entanglement, in the form of quantum correlation between Stokes operators
of two light beams, have been recently demonstrated \cite{Bowen2}. In this work 
the entanglement is of macroscopic nature, and
spatial degrees of freedom do not play any role since the beams are single-mode. 
Continuous variable polarization entanglement which takes into account spatial  spatial degrees of freedom of light beams 
is described in \cite{Zambrini}, where we study the properties of the light emitted 
by a type II optical parametric oscillator
below threshold.\\
The analysis of this paper is rather focussed on providing a bridge between the miscroscopic and macroscopic domain, 
since our  model is able to describe the polarization entanglement in parametric down-conversion
with a continuous passage from the single photon pair production regime to the regime of high down-conversion
efficiency. \\
Besides its fundamental interest, we believe that the form of entanglement described in this work
can be quite  promising 
for new quantum information schemes, due the increased number  of
degrees of freedom in play (photon number, polarization, frequency and spatial degrees
of freedom), and is well inserted in the recent trend toward entangled state of increasing 
complexity (see e.g. \cite{Bowmeester}, where a four-photon polarization entangled state is characterized)

The paper is organised as follow. Section \ref{Sec1} describes the model  for
spontaneous parametric down-conversion, in terms of propagation equations for field operators.
Similar models are known in literature (see e.g. \cite{Misharev} and references quoted therein),
but besides presenting it in a systematic way, we include all the relevant 
features of propagation through a nonlinear crystal and we provide a precise link with 
the empirical parameters of real crystals. Section \ref{Sec2} is devoted to the description
of the quantum polarization properties of the down-converted light.
Stokes operators definition and properties are briefly reviewed in section \ref{secdef}. In Sec \ref{Stokescorr}
we generalize this definition to a local measurement in the beam far field plane, and we introduce the spatial
correlation functions of interest. Analytical and numerical results for the degree of  correlation of the various
Stokes parameters detected from symmetric portions of the beam cross-section are presented
in sections \ref{sec01}, \ref{sec23}, both in the case when a narrow frequency filter is employed (Sec.\ref{secnarrow}),
and when the filter is broad-band (Sec.\ref{broad}).
Section \ref{Sec3} provides an alternative description of the system and of its polarization correlations in the
framework of the quantum state formalism.  Section \ref{conclusions} finally concludes.

\section{A multi mode model for type II parametric down-conversion}
\label{Sec1}
\subsection{Field propagation}
The starting point of our analysis is an equation describing the propagation of 
the three waves (signal, idler and pump)
inside a nonlinear $\chi^{(2)}$ crystal. We consider a crystal slab of length $l_c$, 
ideally infinite in the transverse
directions, cut for type II quasi-collinear 
phase-matching.
In the framework of the slowly varying envelope approximation 
the electric field operator associated to the three waves is described by
means of three quasi-monochromatic wave-packets.
We take the $z$ axis as the laser pump mean propagation 
direction (Fig.1), and indicate with $\vec{x}=(x,y)$
the position coordinates in a generic transverse plane. 
$\hat{E}^{(+)}_i(z,\vec{x},t)$, $i=o,e,p$ designate  the positive frequency part of 
the field operator (with dimensions of a photon annihilation operator)
associated 
to the ordinary  ($i=o$, the "signal")  and 
extraordinary ($i=e$, the "idler") polarization components  of the downconverted beam, 
and ($i=p$ the "pump") the  high frequency laser beam activating the down-conversion process.
Next we introduce their Fourier transform in time and in the transverse domain:
\be
\hat{A}_i (z,\q, \om) = \int \frac{\dd \x}{2\pi} \int \frac{\dd t}{\sqrt{2\pi}} 
		\e^{-\im \q\cdot \x}\e^{\im (\omega_i +\om)t}\hat{E}^{(+)}_i(z,\x,t) 
\qquad i=o,e,p
				\label{envelope}
\ee
Here $\q $ is the transverse component of the wave-vector  and 
$\om$ represents the frequency offset from the carriers $\omega_o+\omega_e=\omega_p$.
In the following, we shall assume degenerate phase matching, so that $\omega_o=\omega_e=
\omega_p/2$.
It is convenient to subtract from the field operators the fast variation along z 
arising from linear
propagation inside the birefringent crystal. We write:
\be
\hat{A}_i (z,\q, \om) =\exp {\left[ \im k_{iz}(\q,\om)z\right]  }  \hat{a}_i(z,\q,\om) 
				\label{interactionp} \: ,
\ee
where $k_{iz}(\q,\om)= \sqrt{k_i^2  ({\q,\omega_i+\om}) -q^2} $ is the projection of the wave-vector
along the z direction, with $k_i(\q,\omega_i+\om)$ being the wave number  of the i-th wave.
In the absence of any nonlinear interaction, we would have
\be 
\frac{\dd}{\dd z} \hat{a}_i(z,\q,\om)=0 \,
\ee
being Eq.(\ref{interactionp}) with
$\hat{a}_i(z,\q,\om)=\hat{a}_i(z=0,\q,\om)$ the forward
solution of Maxwell wave equation in linear dispersive media.  For the pump wave,
we assume that the intense laser pulse is undepleted by the down-convertion process,
so that $\hat{a}_p(z,\q,\om)=\hat{a}_p(z=0,\q,\om) $. Moreover, we assume that the pump is an intense
coherent beam and the operator can be replaced by its classical mean value $\alpha_p(\q,\om)$.\\
For the signal and idler beams, the variation of $\hat{a}_i$ operators along z is only
due  to the nonlinear term, proportional to the $\chi^{(2)}$ material second order susceptibility.
This is usually very small, so that $\hat{a}_i$ are slowly varying along z. This allows us to neglect
the second order derivative with respect to $z$ in the wave-equation. Hence
the resulting propagation equation takes the form (see also \cite{Brambilla02} 
for more details, and \cite{Scotto02} for an alternative derivation):
\be
\frac{\dd }{\dd z} \hat{a}_i \left(z,\q,\om\right) = \chi \int \dd \q^{\,\prime}  \, \int \dd \om^{\prime} \, \,
		\alpha_p\left( \q +\q^{\,\prime}, \om +\om^{\prime} \right) \hat{a}_j^{\dagger} 
			\left(z,\q^{\,\prime},\om^{\prime}\right) \e^{-\im \Delta_{ij} (\q,\q^{\,\prime}; \om, \om^{\prime})z/l_c}
					\qquad i\ne j =o,e
				\label{eqz}  \: ,
\ee
where $\chi$ is a parameter proportional to the second order susceptibility of the medium, and 
\be
\Delta_{ij} \left(\q,\q^{\,\prime}; \om, \om^{\prime} \right)= l_c \left[ k_{iz} \left(\q,\om\right)+   
		 k_{jz} \left(\q^{\,\prime}, \om^\prime \right)-
k_{pz} \left(\q+\q^{\,\prime}, \om+\om^\prime \right) \right]
				\label{mismatch}  
\ee
is the phase mismatch function. Equation (\ref{eqz} ) describes all the possible microscopic processes 
through which a pump photon of frequency $\omega_p + \om +\om'$, propagating in the direction $\q+\q^{\, \prime} $
is annihilated at position $z$ inside the crystal,
and gives rise to a signal and an idler
photon, with frequencies $\omega_p/2 + \om $, $\omega_p/2 + \om' $, and transverse wave vectors $\q$, $\q^{\, \prime} $, 
with an overall conservation of energy and transverse momentum. The effectiveness of each process
is weighted by the phase mismatch function (\ref{mismatch}), which accounts for conservation of the longitudinal momentum. In the limit
of an infinitely long crystal, where longitudinal radiation momentum has to be conserved, only those processes
for which $\Delta_{ij}=0$ are allowed. For a finite crystal, however, the phase matching function has  finite
 bandwidths, say $q_0$
in the transverse domain and $\om_0$ in the frequency domain.\\
Equation (\ref{eqz} ) couples all the signal and idler spatial and temporal frequencies within the angular
bandwith of the pump $\delta q \approx \frac{1}{w_p}$, with $w_p$ being the pump beam waist, and within
the pump temporal spectrum $\delta \om \approx 1/\tau_p$, where $\tau_p$ is the pump pulse duration. In general, 
no analytical solution is available and one has to resort to numerical methods in order  to 
calculate the quantities of interest, as described in\cite{Brambilla02}.\\
A limit where analytical results can be obtained is that of a pump waist and a pump duration large enough, so  that
$\delta q <<q_0$, $\delta \om <<\om_0$. In this case the pump beam can be approximated by a plane wave
\be
\alpha_p \left( \q +\q^{\prime}, \om +\om^{\prime} \right) \to \alpha_p \delta \left( \q +\q^{\prime}\right)
\delta\left(\om +\om^{\prime}  \right) \label{PW}  \: ,
\ee
Equation (\ref{eqz}) reduces to
\beqa
l_c \frac{\dd }{\dd z} a_o \left(z,\q,\om\right) &=& \sigma 
		a_e^{\dagger} 
			\left(z,-\q,-\om \right) \e^{-\im \Delta (\q, \om)z/l_c}
					\nonumber \: ,\\
l_c \frac{\dd }{\dd z} a_e \left(z,-\q,-\om\right) &=& \sigma 
		a_o^{\dagger} 
			\left(z,\q,\om \right) \e^{-\im \Delta (\q, \om)z/l_c}
				\label{PWeqz}  \: ,
\eeqa
where $\sigma=l_c \chi \alpha_p$ is a linear gain parameter, and
\be
\Delta(\q, \om)= l_c \left[ k_{oz}(\q,\om) +k_{ez}(-\q,-\om)-k_p \right]\: .
					\label{Delta}  
\ee
is the phase mismatch of a couple of  ordinary and extraordinary waves propagating with symmetric 
transverse wave vectors $\q$ and $-\q$, and with frequencies $\omega_p/2 +\om$, $\omega_p/2 -\om$. 

Solution of the propagation equation (\ref{PWeqz})   is found  in terms of 
the  field distributions
at the input face of the crystal. 
 Coming back  to the  field operators
defined by Eq. (\ref{envelope}), we define the field operators at the input and output faces of the crystal slab as
\beqa
\hat{A}_i^{in}(\q,\om)&=& \hat{a}_i (z=0,\q,\om) \\
\hat{A}_i^{out}(\q,\om)&=& \hat{a}_i (z=l_c,\q,\om)\exp{\left[ \im  k_{iz}(\q,\om )l_c\right]}  
\eeqa
By solving Eq.(\ref{PWeqz}) the transformation from the input to the output operators is found in the form of 
a two-mode squeezing transformation:
\beqa
\hat{A}_o^{out}(\vec{q},\Omega)&=& U_o(\q,\Omega) \hat{A}_o^{in}(\vec{q},\Omega) +
		V_o(\q,\Omega) \hat{A}_e^{\dagger \, in}(-\vec{q},-\Omega) \nonumber\\
\hat{A}_e^{out}(\vec{q},\Omega)&=& U_e(\q,\Omega) \hat{A}_e^{in}(\vec{q},\Omega) +
		V_e(\q,\Omega) \hat{A}_o^{\dagger \, in}(-\vec{q},-\Omega) 
						\label{inout1} \; ,
\eeqa
linking only symmetric modes $\q,\om$ and $-\q ,-\om$ in the signal and idler beams (see e.g. \cite{Misharev} 
for a similar transformation in the type I case).
If we require that free space commutation relations
\be
\left[ \hat{A}_i^{in}(\q,\om), \, \hat{A}_j^{in}( \q^{\, \prime},\om^{\prime})\right]= \delta_{i,j} \delta(\q-\q^{\, \prime}) 
				\delta (\om - \om ^\prime) \qquad i,j=o,e
										\label{commutator}
\ee
are preserved from the input to the output, 
it can be easily shown that the complex coefficients of the transformation (\ref{inout1}) need to 
satisfy the following conditions:
\beqa
\left|U_i (\q,\Omega)\right|^2-\left|V_i (\q,\Omega)\right|^2&=&1 \qquad (i=o,e)
					\label{uni1} \\
U_o (\q,\Omega)V_e(-\q,-\Omega) &=& V_o (\q, \Omega) U_e (-\q, -\Omega)
				\label{uni2}
\eeqa
By taking the modulus of the second relation and making use of the first two ones,
the complex equation (\ref{uni2}) can be written as two equivalent real equations:
\beqa
\left|V_o (\q,\Omega)\right|^2 &=& \left|V_e (-\q,-\Omega)\right|^2 \label{uni3} \\
\mbox{arg} \left[ U_o (\q,\Omega)V_e(-\q,-\Omega)\right] &=&
\mbox{arg} \left[ V_o (\q,\Omega)U_e(-\q,-\Omega)\right]  := 2 \psi(\q,\Omega)
\eeqa
With this in mind,  the coefficients of the transformation (\ref{inout1}) can be recasted
in the form
\be
\begin{array} {ll}
U_o (\q,\Omega)= U(\q,\Omega) e^{i \varphi(\q,\Omega)} \quad , \quad 
&V_o (\q,\Omega)=  V(\q,\Omega) e^{i \varphi(\q,\Omega)} 
			\\
U_e (\q,\Omega)= U(-\q,-\Omega) e^{-i \varphi(-\q,-\Omega)} \quad , \quad
&V_e (\q,\Omega)= V(-\q,-\Omega) e^{-i \varphi(-\q,-\Omega)} \; ,
\end{array}
			\label{bla2}
\ee
with
\beqa
U(\q, \Omega) &=& \cosh r(\q,\Omega) e^{i \psi(\q,\Omega)} e^{i \theta(\q,\Omega)} \nonumber \\
V(\q,\Omega)&=& \sinh r(\q,\Omega)e^{i \psi(\q, \Omega)} e^{-i \theta(\q,\Omega)},
				\label{bla3}
\eeqa
where $r(\q,\Omega)$,$\varphi(\q,\Omega$, 
$\psi(\q,\Omega$,  $\theta(\q,\Omega$, are independent real functions of $\q,\Omega$.

We outline that  the form of the transformation (\ref{inout1}), together with the unitarity requirements
(\ref{uni1}, \ref{uni2}), are enough the derive the general form of the results presented in this paper.
In the following, we shall present results for a specific device, namely travelling 
wave parametric down-conversion,
and we shall take as an example the case of a BBO ($beta$-barium-borate) crystal. However, 
a similar  investigation can be carried out
for any device characterized by an input/ouput transformation of the form (\ref{inout1}). 
The case of Type II parametric
down-conversion inside an optical resonator is for example investigated in \cite{Zambrini}.

More insight into the problem is gained by 
looking at the explicit solution of the propagation equation (\ref{PWeqz}). We obtain
\beqa
U(\q,\Omega)&=& \e^{\im \frac{k_p l_c}{2} }
	\left\{  \cosh \left[\Gamma (\q,\Omega) \right] + 
	\im \frac{\Delta (\q,\Omega)}{2\Gamma (\q,\Omega)}          
\sinh \left[\Gamma (\q,\Omega) \right]	\right\} \: ,
						\label{U}\\
V(\q,\Omega)&=& \e^{\im \frac{k_p l_c}{2} }
\frac{\sigma}{\Gamma (\q,\Omega)} \sinh \left[ \Gamma (\q,\Omega) \right]
						\label{V}\: ,\\
\varphi (\q,\om) &=& \frac{l_c}{2} \left[  k_{oz}(\q,\om) - k_{ez}(-\q, -\om)\right] \: ,
				\label{phi}
\eeqa 
with
\be
\Gamma (\q,\Omega)=\sqrt{ \sigma^2 - \frac{\Delta^2(\q,\om)}{4}} \: ,
			\label{Gamma}
\ee
and $\Delta(\q,\om) $ is the phase mismatch function defined by (\ref{Delta}). 

\subsection{Phase matching curves}

The gain functions (\ref{U}, \ref{V}) reach  their maximum value for phase matched modes,
that is, the  modes for which
$\Delta(\q,\om)=0$. By assuming the validity of the paraxial and slowly varying envelope approximations,
the longitudinal wave-vector components $k_{iz} (\q, \om)$ can be expanded in power series of $\q, \om$.
By  keeping only the leading terms we obtain:
\be
k_{iz}(\q,\om) \approx k_{i}  + \frac{1}{v_g^i}  \om
+ \frac{1}{2}\frac{\dd^2 k_{i}}{\dd \om^2}  \om^2 + \frac{\dd k_{i}}{\dd q_y}  q_y
-\frac{q^2}{2k_i} \qquad i=o,e\: .
				\label{exp}
\ee
The first term at r.h.s. is 
$k_i= n_i (\omega_i, \q=0) \, \omega_i /c$, with 
$n_i$  being the index of refraction at the carrier frequency of 
an ordinary (extraordinary)
wave propagating along $z$ direction. The  second term,
$ \frac{1}{v_g^i} \om=\frac{\dd k_{i}}{\dd \om} \om $, accounts for the fact that 
the three wave-packets move with different group velocities $v_g^i$. The third
term describes the effects of temporal dispersion. In writing the fourth term,
we assumed that the crystal is uniaxial and the crystal optical axis lies in the z-y
plane. This term is present only for the extraordinary waves , and
$\frac{\dd k_{i}}{\dd q_y}= -\rho_i$ where $\rho_i$ is the walk-off angle of the wave.
Finally, the last term describes the effects of diffraction for a paraxial wave.\\
With this in mind, the phase matching function can be written in the form:
\be
\Delta(\q,\om)  = \Delta_0  + \rho_e l_c q_y -\frac{q^2}{q_0^2} + \om \tau_{coh} +\frac{1}{2} 
\epsilon (\om \tau_{coh})^2
			\label{delta}
\ee
 where
\begin{description}
\item{1)}
\be
\Delta_0  = (k_o +k_e -k_p) l_c  
			\label{delta0}
\ee
is the collinear phase mismatch (i.e. the phase mismatch of the three waves 
at the carrier frequencies when propagating along the longitudinal direction);
\item{2)}
\be
q_0= \sqrt{  \frac{1}{l_c}   \frac{k_e +k_o}{2k_e k_o}  }= 
		\sqrt{ \frac{2\pi}{\lambda l_c}   \frac{n_e +n_o}{2n_e n_o}  }
					\label{q0}
\ee
with $\lambda=4 \pi c / \omega_p$ being the wavelength in vacuum at the carrier frequency $\omega_p/2$, and $n_e, n_o$ the odinary
and extraordinary refraction indexes inside the crystal at the carrier frequency. This parameter defines
the typical bandwidth of phase matching in the transverse q-space domain. Its inverse $l_{coh}=1/q_0$ will be referred 
to as the {\em coherence length}. 
\item{3)}
\be
\tau_{coh}= \frac{l_c}{v_g^o} -\frac{l_c}{v_g^e} 
					\label{tauc}
\ee
with $v_g^i$ being the group velocities of the two waves, is the the difference between the time taken by
the signal and idler wave-packets to cross the crystal. This defines the typical scale of variation of gain functions
in the temporal domain for type II phase matching, and it will be referred to as the amplifier {\em coherence time}.
\item{4)} 
Finally
\be
\epsilon= \left(\frac{\dd^2 k_o}{\dd \om^2}+\frac{\dd^2 k_e}{\dd \om^2} \right) \frac{l_c}{\tau_{coh}^2}
\ee
is a dimensioneless parameter that depends on the temporal dispersion properties of the signal and idler pulses 
(typically $\epsilon << 1$).
\end{description}
\begin{figure}[ht]
\centerline{\scalebox{.60}{\includegraphics*{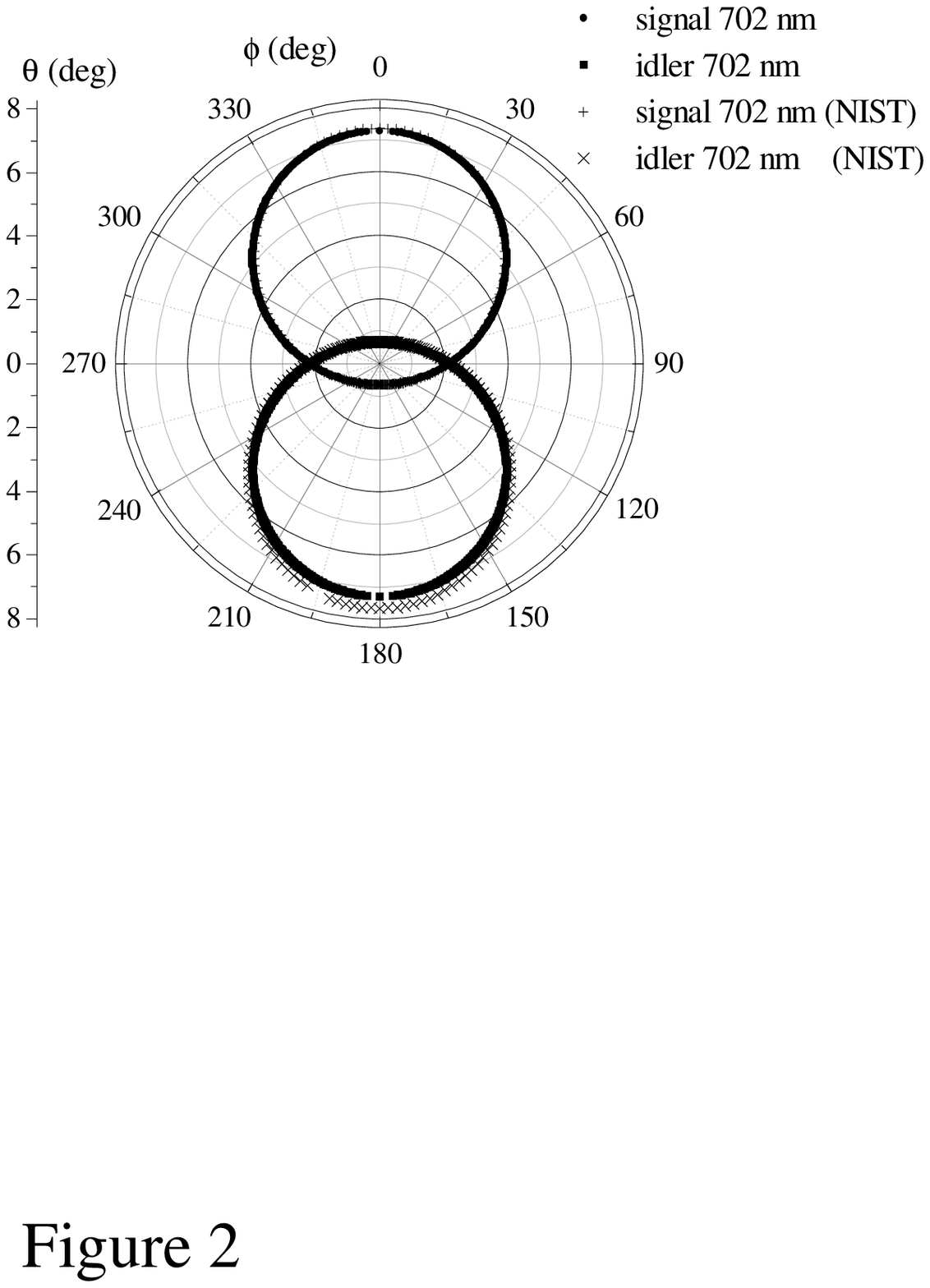}}}
\caption{Polar plot of phase matching curves in a BBO crystal. Comparison of the approximated 
formula (\ref{delta}) with the ``exact" phase matching formula, calculated with the method in \cite{NIST} (NIST).
$\theta$ is the polar angle from the pump direction of propagation, $\phi$ is the azymuthal angle.
$\lambda_{signal}=\lambda_{idler}=702 nm $. ($\lambda_{pump}=351 nm$), and the pump propagates at an angle of $49.6$ 
degrees from the crystal optical axis}
\label{fig2}
\end{figure}
The equation $\Delta (\q,\om)=0$ defines in the $(q_x, q_y)$ plane a circonference, centered at the position
\be
q_x=0, \quad q_y=q_C = \frac{1}{2} q_0^2 \rho_e l_c  
\ee 
and with  radius given by  
\be
q_R=q_0 \sqrt{\Delta_0  + \frac{q_C^2}{q_0^2}+ \om \tau_{coh} + \frac{1}{2} \epsilon (\om \tau_{coh})^2 } \: .
\ee
This corresponds to the phase matched modes for the signal (ordinary) wave. Phase matched modes for the 
idler wave, emitted at the frequency $-\om$, lye on the symmetric circonference.\\
Figure \ref{fig2} plots some examples of this phase matching circles, in the form of polar plot, with $\theta$ being
the polar angle from the pump direction ($z$ axis) outside the crystal and $\phi$ the azymutal angle around $z$.
Parameters are those of a BBO crystal, cut for degenerate phase matching at $49.6$ degrees,
for a pump wavelength of $351$ nm. They have been calculated with the help of empirical Sellmeier
formulas for refraction indices in Ref.\cite{Sellmeier}. 
For comparison, superimposed to the curves calculated by means of Eq.(\ref{delta}),
the figure shows the ``exact" phase matching curves, calculated with the method described in \cite{NIST},
by means of a public domain numerical routine available at \cite{NISTweb}. The plots show a rather good agreement,
in any case within the error inplicit in the use of empirical Sellmeier formulas.
\section{Polarization correlation: quantum field formalism}
\label{Sec2}
\subsection{Stokes operators:definition and properties}
\label{secdef}
Quantum-optical polarization properties of light  are conveniently 
described within the formalism of Stokes operators, 
which represent the quantum conterparts of the Stokes vectors of classical optics. 
The polarization state of a classical beam can be described by means of a Stokes vector, and of its associated 
Poincar\`e sphere. Stokes vectors pointing on the equator
of the sphere represents linearly polarized light; if the vector points in the 
positive (negative) directon of  $S_1$ 
the light is horizontally (vertically) polarized, while the $S_2$ direction identifies light polarized at 
45 (-45) degrees. The $S_3$ direction corresponds to circulary (right and left) polarized light. A fourth parameter,
$S_0$ is the total beam intensity, and gives the radius of the Poincar\`e sphere.
 For a polarized beam $S_1^2+S_2^2+S_3^2=S_0^2$, so that the polarization state is represented by 
a point on the sphere surface.\\
In the quantum mechanical description of light polarization, Stokes parameters are replaced by a set 
of four Stokes operators.
For a single mode of an electromagnetic field, they are defined in terms 
of the photon annihilation operators for a vertical and horizontal linear polarization mode $\hat{a}_H, \hat{a}_V$, as:
\beqa
\hat S_0&=& \hat{a}^\dagger_H \hat{a}_H + \hat{a}^\dagger_V \hat{a}_V \\
\hat S_1&=&  \hat{a}^\dagger_H \hat{a}_H - \hat{a}^\dagger_V \hat{a}_V \\
\hat S_2&=&  \hat{a}^\dagger_H \hat{a}_V + \hat{a}^\dagger_V \hat{a}_H = \hat{a}^\dagger_{45} \hat{a}_{45} - \hat{a}^\dagger_{-45}  \hat{a}_{-45}  \\
\hat S_3&=&  -\im \left( \hat{a}^\dagger_H \hat{a}_V - \hat{a}^\dagger_V \hat{a}_H \right)=\hat{a}^\dagger_{R} \hat{a}_{R} - \hat{a}^\dagger_{L}  \hat{a}_{L} \: , 
\eeqa
where $\hat{a}_{45}$ , $\hat{a}_{-45}$ denote annihilation operators on the oblique polarization basis, and 
$a_R$ $a_L$ are annihilation operators on the circular right and circular left polarization basis.
The first two operators represent, respectively, the sum and difference of photon numbers in the vertical/horizontal 
polarization basis. Operators $\hat S_2$ and $\hat S_3$ are the difference of photon numbers in the oblique and circular
polarization basis, respectively. All these observables can be  measured  by means of a polarizing beam splitter
and quarter and half wave plates, as e.g. described in \cite{BornWolf}. However, while operator $\hat S_0$ commutes
with all the others, the remaining three do not:
\be
[\hat S_1, \hat S_2]= 2\im \hat S_3\: , \quad [\hat S_2, \hat S_3]= 2\im \hat S_1 \: ,\quad 
[\hat S_3, \hat S_1]= 2\im \hat S_2 \; .
\ee
The set of Stokes operators has angular momentum-like commutation relation, and the associated observables are in general 
non compatible. The quantum state of a light beam 
cannot be any more visualized as a point on the Poincar\'e sphere, since quantum noise introduces 
a minimum uncertainty in the values of the Stokes parameter. Polarization squeezed states,
whose uncertainty can be represented by an ellipsoid (see e.g. \cite{Korolkova}) has been recently realized
\cite{Bowen}.

\subsection{Stokes operator correlation in the far field of parametric down-conversion}
\label{Stokescorr}
\begin{figure}[h]
\centerline{\scalebox{.50}{\includegraphics*{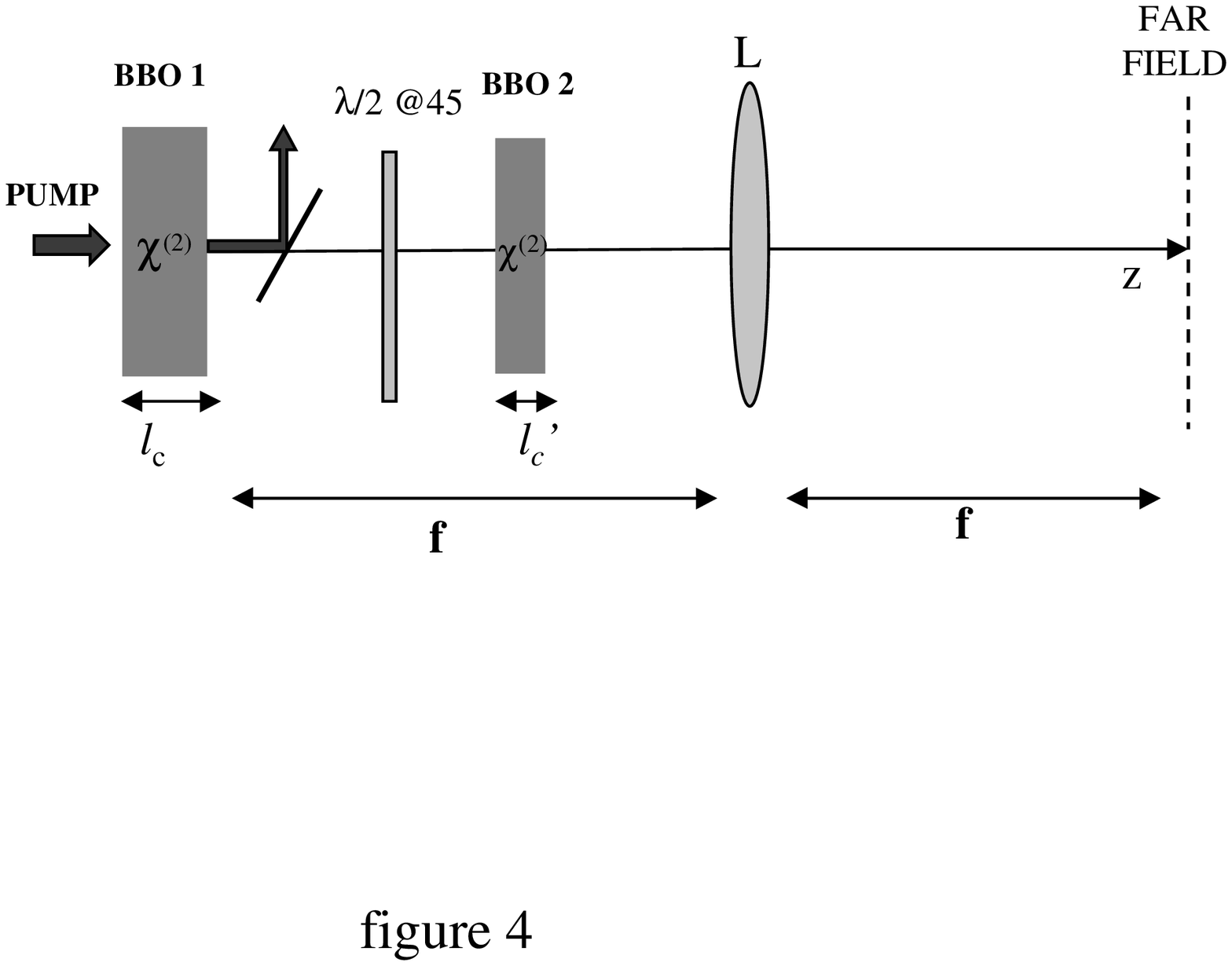}}}
\caption{Schematic set-up for a measurement in the far field plane with a compensation crystal. BBO1, down-conversion
crystal of length $l_c$; $\lambda/2 @ 45$ half-wave plate, rotates polarization by 90 degrees; BBO2, compensation crystal of length
$l_c^\prime$, $L$ lens of focal length $f$. }
\label{figsetup}
\end{figure}

The main idea of this paper  is to study the quantum correlation
between  Stokes operators measured from symmetric portions of the far field beam cross-section. To this 
end, we consider a measurement of the Stokes 
operators over
a small region
$D(\x)$ centered around a position $\x$ in the far-field plane of the down-converted field,
and over a detection
time $T$ (tpically we will take $T$ much larger than the crystal coherence time).
\be
\hat{S}_{i}(\x) = \int_{T} dt^{\prime} \int_{D(\x)}d\x ^{\,\prime} \hat{\sigma}_i (\x^{\,\prime} ,t^{\prime} ) \; ,
			\label{mystokes}
\ee
where
\beqa
 \hat{\sigma}_0(\x ,t) &=& \hat{A}_{o}^{\dagger} (\x ,t) \hat{A}_{e} (\x ,t) + \hat{A}_{o}^{\dagger}(\x,t) \hat{A}_{e}(\x,t) \; , \label{sigma0} \\
\hat{\sigma}_{1}(\x,t) &=& 
\hat{A}_{o}^{\dagger}(\x,t) \hat{A}_{o}(\x,t) - \hat{A}_{e}^{\dagger}(\x,t) \hat{A}_{e}(\x,t) \:, \label{sigma1} \\
\hat{\sigma}_{2}(\x,t) &=& \hat{A}_{o}^{\dagger}(\x,t) \hat{A}_{e}(\x,t) + \hat{A}_{e}^{\dagger}(\x,t) \hat{A}_{o}(\x,t) \; , \label{sigma2}  \\
\hat{\sigma}_{3}(\x,t) &=& -{\rm i }\left[
\hat{A}_{o}^{\dagger}(\x,t) \hat{A}_{e}(\x,t) - \hat{A}_{o}^{\dagger}(\x,t) \hat{A}_{e}(\x,t) \right] \label{sigma3}  \: .
                       \label{Stokes}
\eeqa
$\hat{A}_{o/e}$ denotes the field operator for the ordinary/extraordinary polarized beam 
in the far-field plane, which can be observed in the focal plane of a lens, placed as 
shown in figure \ref{figsetup}.\\
By using the free field commutation relations (\ref{commutator}), it can be easily shown that
\be
[\hat S_1(\x), \hat S_2(\x)]= 2\im \hat S_3 (\x) \: , \quad [\hat S_2(\x) , \hat S_3(\x) ]= 2\im \hat S_1(\x)  \: ,\quad 
[\hat S_3(\x) , \hat S_1(\x) ]= 2\im \hat S_2 (\x)\; ,
\ee
while operators measured from different (and not connected) detection pixels commute.

In the following, we shall consider Stokes operator correlation functions of the form:
\begin{equation}
\langle \delta \hat S_i (\x) \, \delta \hat S_j (\x^{\, \prime}) \rangle 
= 
\langle \hat S_i (\x) \, \hat S_j (\x^{\, \prime}) \rangle - 
\langle \hat S_i (\x)\rangle  \, \langle \, \hat S_j (\x^{\, \prime}) \rangle \: , \qquad (i,j=0, \ldots 3) \: .
					\label{Sij}
\end{equation}
A useful tool for calculation  are  the 
correlation functions of the Stokes operator densities (\ref{sigma0}-\ref{sigma3}) : 
\begin{equation}
G_{ij} (\x, \x^{\, \prime}; \tau) = 
\langle \hat \sigma_i (\x, t+\tau) \, \hat \sigma_j (\x^{\, \prime}, t) \rangle - 
\langle \hat \sigma_i (\x, t+\tau)\rangle  \, \langle \, \hat \sigma_j (\x^{\, \prime},t) \rangle 
					\label{Gij}
\end{equation}
and their spectral densities
\begin{equation}
\tilde G_{ij} (\x, \x^{\, \prime}; \om) = 
\int d \tau \e^{\im \om \tau } G_{ij} (\x, \x^{\, \prime}; \tau)
					\label{tildeGij}
\end{equation}
Their relation with the correlation functions of the measured Stokes operator (\ref{Sij}) is given by: 
\begin{equation}
 \langle \delta \hat S_i (\x) \, \delta \hat S_j (\x^{\, \prime}) \rangle = 
\int_{D(\x)} d{\x}_1 \int_{D(\x^{\, \prime})} d{\x}_1^{\, \prime}
\int d \om  \frac{T^2}{2\pi} {\rm sinc}^2 \left( \om  \frac{T}{2} \right)
\tilde G_{ij} \left( \x_1, \x_1^{\, \prime}  ; \om \right) \: .
					\label{Gg}
\end{equation}
We notice that:
\be
\lim_{T \to \infty}   \frac{T}{2\pi} {\rm sinc}^2 \left( \om  \frac{T}{2} \right)  = \delta(\om) \; ,
\ee
and that this function acts under the integral as a frequency filter with bandwidth $\Delta \om= 2\pi/T$.
We shall assume in the following that the detection time is much larger than the crystal coherence time.
Under this assumption
\begin{equation}
 \langle \delta \hat S_i (\x) \, \delta \hat S_j (\x^{\, \prime}) \rangle 
\stackrel{ T \gg \tau_{coh}} {\longrightarrow}  T \int_{D(\x)} d{\x}_1 \int_{D(\x^{\, \prime})} d{\x}_1^{\, \prime}
\tilde G_{ij} \left( \x_1, \x_1^{\, \prime}  ; \om=0 \right)
					\label{limitG}
\end{equation}

When a lens of focal length $f$ is placed at a focal distance both from the crystal output
plane and the observation plane (see figure \ref{figsetup}), 
the field operators in the far-field plane are connected to those at the crystal output 
by the usual mapping \cite{Goodman}
\be
\hat{A}_i (\x,\om)= \frac{2\pi}{\im \lambda_i f} \hat{A}_i^{out} \left( \q=\frac{2\pi \x}{\lambda_i f}, \om\right) \; ,
			\label{near-far}
\ee
where $f$ is the focal length of the lens used to image the far field plane and $\lambda_i$ is the
wavelength (in vacuum) at the frequency $\omega_p/2 +\om$.\\
Since the light statistics is Gaussian (the output operators are 
obtained by a linear transformation acting on input vacuum field operators)
all expectation values and correlation functions of interest can be calculated by making 
use of the second order moments of field operators.
These can be easily calculated in the far field plane by assuming 
that the down-converted field operators at the input crystal face are in the vacuum state
and by using the input/output relations (\ref{inout1}), toghether with (\ref{near-far}), thus obtaining
\beqa
\left\langle \hat{A}_i^\dagger (\x, \om) \hat{A}_j ( \x^{\,\prime}, \om^{\,\prime} )  \right\rangle &=& \delta_{i,j}
\delta(\x -\x^{\,\prime}) \, \delta(\om -\om^{\,\prime} )\left| \V_i(\x, \om)\right|^2 \label{aadaga} \: , \\
\left\langle  \hat{A}_i(\x, \om) \hat{A}_j ( \x^{\,\prime}, \om^{\,\prime} )  \right\rangle &=& -\frac{\lambda_j}{\lambda_i}  
(1-\delta_{i,j})
\delta(\x  \frac{\lambda_j}{\lambda_i}  +\x^{\,\prime}) \, \delta(\om +\om^{\,\prime} )
\U_i(\x,\om)  \V_j (-\x \frac{\lambda_j}{\lambda_i}, -\om)
		 \label{aa} \: , \qquad (i,j=o,e) \; .
\eeqa
In this formula
\be
\U_i (\x,\om)= U_i \left(\q= \x \frac{2\pi}{\lambda_i f}, \om \right)\: , \qquad 
\V_i (\x,\om)= V_i \left(\q= \x \frac{2\pi}{\lambda_i f}, \om \right) 
				\label{baruv}
\ee
where $U_i, V_i$ are  are the gain functions defined by (\ref{bla2}--\ref{Gamma}).
It can be noticed the presence of the  nonzero ``anomalous" propagator (\ref{aa}), a term which
is characteristic of processes where particles are created in pairs.
In order to simplify the notation, in the following we shall consider the case $\om << \omega_p/2$,
and take $\lambda_o=\lambda_e=\lambda=2 \lambda_p$. In a real experimental implementation, however,
the validity of such an approximation should be carefully checked when not using narrow frequency filters;
 twin photons produced at different wavelengths $\lambda_e$, $\lambda_o$, and 
travelling with symmetric $\q$, $-\q$ transverse wave vectors are actually propagating at different angles
from the pump and will  be intercepted in the far field at two sligtly different radial positions.\\
The fact that the field spatial correlation are perfectly localized in the far field (the Dirac-delta
form of the correlation peak) is a consequence of the traslational symmetry of the model in the transverse plane
(plane wave pump and a  crystal slab infinite in the transverse direction). A trivial formal fault 
is that the far field mean intensity of the downconverted beams diverges, as a consequence of the infinite 
energy of a plane-wave pump.
This artificial divergence can be  formally eliminated with 
the trick used  in \cite{Kolobov99,Brambilla}, where a finite size pupil was inserted at the output
face of the crystal. The spatial Dirac-delta functions in Eqs.(\ref{aadaga},\ref{aa}) are substituted
by a finite version, and a typical resolution area, proportional to the
diffraction spot
of the pupil in the far field plane, is introduced in the scheme. 
For a pupil of transverse area $S_P$, this  is given by $D_R=(\lambda f)^2/S_P$. The typical scale of variation
of the gain functions (\ref{baruv}) in the far field plane is 
\be
X_0=q_0 \lambda f/(2 \pi) \: ;
					\label{X0}
\ee
 when $X_0$ is much larger
than the resolution area (or, equivalently, when the pupil size is much larger than the amplifier coherence length),
the mean photon number distribution in the far field plane is given by
\beqa
\langle\hat N_i (\x) \rangle  &=& \int_{D(\x)} d\x^{\, \prime} \int_{T} d t 
	\langle \hat{A}^\dagger_i (\x^{\,\prime} ,t) \hat{A}_i (\x^{\, \prime} ,t) \rangle = \nonumber \\
&\approx& \frac{T}{D_R} \int_{D(\x)} d\x^{\, \prime}  \int \frac{d \om}{2\pi} \left| \V_i (\xp,\om)\right|^2 
			\label{meanN} \: ,
\eeqa
When the finite size of the pump is taken into account in a numerical model~\cite{Brambilla02}, it is easily seen that the resolution
area is rather given in terms of the spot size of the pump as it is imaged in the far field plane. For a Gaussian pump
of waist $w_p$, 
$D_R \approx (\lambda f)^2 / (\pi w_p^2)$. 

In this limit of small resolution area, the mean  value of Stokes operators is given by:
\beqa
\langle \hat S_0 (\x) \rangle &=& \frac{T}{D_R} \int_{D(\x)} d\x^{\, \prime}  \int \frac{d \om}{2\pi} 
\left[ \left| \V_o (\xp,\om)\right|^2 +\left| \V_e (\xp,\om)\right|^2 \right] 
						\label{S0mean}\\
\langle \hat S_1 (\x) \rangle &=& \frac{T}{D_R} \int_{D(\x)} d\x^{\, \prime}  \int \frac{d \om}{2\pi} 
\left[ \left| \V_o (\xp,\om)\right|^2 - \left| \V_e (\xp,\om)\right|^2 \right] 
						\label{S1mean} \\
\langle \hat S_2 (\x) \rangle &=& \langle \hat S_3 (\x) \rangle =0
						\label{S23mean}
\eeqa
\subsection{Correlation in Stokes operators $S_1,S_0$}  \label{sec01}
The first and second Stokes operators represent the sum and the difference, respectively, between the number of ordinary
and extraordinary photons (say horizontally and vertically polarized photons) 
measured from a detection pixel in the far field plane. 
\beqa
\hat S_0 (\x) &=&  \hat N_o (\x) + \hat N_e(\x) \; \\
\hat S_1 (\x) &=& \hat N_o (\x) - \hat N_e(\x)
\eeqa
The plane wave pump model predicts that the number of ordinary and extraordinary photons
collected from any two
symmetric portions of the far field plane are perfectly correlated observables \cite{Brambilla02,Brambilla}.
This result is a direct  consequence of pairwise emission of photons
with horizontal (ordinary) and vertical (extraordinary) polarizations, 
propagating in symmetric directions, as required by transverse 
light momentum conservation. Hence, this model predicts
an ideally perfect correlation, both between $\hat{S}_0 (\x)$,
$\hat{S}_0 (-\x)$, and between $\hat{S}_1 (\x)$, $-\hat{S}_1 (-\x)$ for any choice of the
position $\x$ in the far field (notice that $\hat{S}_0 (\x)$ commutes with
$\hat{S}_1 (\x ^\prime)$). \\
In a more sofisticated numerical model~\cite{Brambilla02}, it is readily seen that
the finite width of the pump profile introduces an uncertainty in the directions 
of propagation of the down-converted photons. As described by the propagation equation (\ref{eqz}),
when a $o$ photon is emitted in direction $\q$, its twin $e$ photon is emitted in the direction
$-\q$ within an uncertainty $\delta q \propto 2/w_p$, which is the bandwidth of the pump
spatial Fourier transform.  A photon number correlation well
beyond the shot noise level is recovered when photons are collected from regions larger
than a resolution area $D_R \approx \pi \left( \delta q \frac{\lambda f}{2 \pi}\right)^2 = (\lambda f)^2 / (\pi w_p^2)$. 

In the limit of a small resolution area, long but straigthforward calculations \cite{calc}
show that:
\beqa
\tilde G_{00} (\x, \xp ; \om) &=& \frac{1}{D_R}\left[ \delta(\x-\xp)F_1(\x,\om) + \delta(\x +\xp) F_2(\x, \om)\right]
				\label{G00} \\
\tilde G_{11} (\x, \xp ; \om) &=& \frac{1}{D_R}\left[ \delta(\x-\xp)F_1(\x,\om) -\delta(\x +\xp) F_2(\x, \om)\right]
				\label{G11} 
\eeqa
with
\beqa
F_1 (\x, \om) & =& \int \frac{d \omega}{2\pi} \left\{  \left| \V_o(\x,\omega)\U_o(\x,\omega +\om)\right|^2
+\left| \V_e(\x,\omega)\U_e(\x,\omega +\om)\right|^2\right\}
										\label{F1}\\
F_2(\x,\om)&=& \int \frac{d \omega}{2\pi} \left\{  \U_o(\x,\omega) \U_o^*(\x,\omega -\om) 
\V_e(-\x,-\omega) \V_e^* (-\x, -\omega + \om)  \right. \nonumber \\
&+& \left. \U_e(\x,\omega) \U_e^*(\x,\omega -\om) 
\V_o(-\x,-\omega) \V_0^* (-\x, -\omega + \om)  \right\} 
										\label{F2}
\eeqa
The correlation functions have two peaks; the first one, located at $\xp=\x$, accounts for the noise in the 
measurement of Stokes
parameter from a single pixel. The  second one is located at $\xp= -\x$ and accounts for correlation (anticorrelation)
between measurements performed over symmetric pixels. By taking into account the unitarity relations (\ref{uni1},\ref{uni3}), 
it can be immediately noticed that when $\om=0$ (corresponding to long detection times) $ F_1(\x,0) =F_2(\x,0)$,
and the two corelation function peaks have the same size. This represents the maximum amount of correlation allowed 
by Schwarz inequality, which requires that
\be
\left|\langle \delta \hat S_i(\x) \, \delta \hat S_i (-\x) \rangle \right| \le 
\left[ \langle \delta \hat S_i (\x) \, \delta \hat S_i (\x) \rangle 
\langle \delta \hat S_i (-\x) \, \delta \hat S_i (-\x) \rangle \right]^{\frac{1}{2}}
\ee
In our case, assuming two symmetric detection pixels $D(\x)$ and $D(-\x)$, we have e.g. 
\beqa
\langle \delta \hat S_1 (\x) \, \delta \hat S_1 (\x) \rangle &=& 
\langle \delta \hat S_1 (-\x) \, \delta \hat S_1 (-\x) \rangle =
\frac{T}{D_R} \int_{D(\x)} d\xp F_1(\xp,0)  \\
\langle \delta \hat S_1 (\x) \, \delta \hat S_1 (-\x) \rangle &=& -\frac{T}{D_R} \int_{D(\x)} d\xp  F_2(\xp,0) = -
\langle \delta \hat S_1 (\x) \, \delta \hat S_1 (\x) \rangle 
				\label{equalpeaks}
\eeqa
Finally, the existence of such a perfect correlation implies that both $\hat S_1 (\x) + \hat S_1 (-\x)$ and
$\hat S_0 (\x) - \hat S_0 (-\x)$ are noiseless observables. For example:
\beqa
\langle \left[\delta \hat S_1(\x) + \delta \hat S_1 (-x) \right]^2\rangle &=& 
2\left[ \langle \delta \hat S_1 (\x) \, \delta \hat S_1 (\x) \rangle +
\langle \delta \hat S_1 (\x) \, \delta \hat S_1 (-\x) \rangle \right]=0
													\label{nonoise}
\eeqa
\subsection{Correlation in Stokes operators $S_2,S_3$}  \label{sec23}
Quite different is the situation for the other two Stokes operators $S_2,S_3$, which involve
measurements of the photon number in a polarization basis 
different from the ordinary and extraordinary ones of the crystal, 
namely in the oblique and circular polarization basis.\\
Calculations along the same lines of those performed for the first two Stokes operators show that
also in this case the correlation functions display two peaks, one representing 
the noise associated to the measurement over a single pixel, the other the correlation between symmetric pixels.
\beqa
\tilde G_{22} (\x, \xp ; \om) &=& \tilde G_{33} (\x, \xp ; \om) 
				\label{G22G33} \\
&=& \frac{1}{D_R}\left[ \delta(\x-\xp)H_1(\x,\om) + \delta(\x +\xp) H_2(\x, \om)\right]
				\label{G22} 
\eeqa
with
\beqa
H_1 (\x, \om) & =& \int \frac{d \omega}{2\pi} \left\{  \left| \V_o(\x,\omega)\U_e(\x,\omega +\om)\right|^2
+\left| \V_e(\x,\omega)\U_o(\x,\omega +\om)\right|^2\right\}
										\label{H1}\\
H_2(\x,\om)&=& \int \frac{d \omega}{2\pi} \left\{  \U_o^*(\x,\omega) \U_e(\x,\omega +\om) 
\V_e^*(-\x,-\omega) \V_o (-\x, -\omega -\om)  \right. \nonumber \\
&+& \left. \U_e^*(\x,\omega) \U_o(\x,\omega +\om) 
\V_o^*(-\x,-\omega) \V_e (-\x, -\omega - \om)  \right\} 
										\label{H2}
\eeqa
However, at difference with the previous case, the two peaks in general do not have the same size, even
for a long measurement time. Letting $\om=0$ in Eqs. (\ref{H1}, \ref{H2}) and using the definition (\ref{bla2}), which is
a consequence of unitarity, we have
\beqa
H_1 (\x, 0) & =& \int \frac{d \omega}{2\pi} \left\{  \left| \V(\x,\omega)\U(-\x,-\omega )\right|^2
+\left| \V(-\x,-\omega)\U(\x,\omega)\right|^2\right\}
										\label{H10}\\
H_2(\x,0)&=& \int \frac{d \omega}{2\pi} 2{\rm Re}\left\{  \U^*(\x,\omega) \U(-\x,-\omega ) 
\V^*(\x,\omega) \V (-\x, -\omega ) \right\} 
										\label{H20} \; , \\
H_1 (\x,0) -H_2 (\x,0) &=& \int \frac{d \omega}{2\pi} \left|   \V^*(\x,\omega) \U (-\x,-\omega) -
\V^*(-\x,-\omega) \U(\x,\omega) \right|^2  
								\label{H10-H20} 
\eeqa
where $\U, \V$ appearing in these equations  are the functions defined by Eqs.(\ref{U},\ref{V}),
calculated at $\q=\x  2 \pi /(\lambda f)$. Moreover,
\beqa
\langle \left[\hat S_2(\x) -\hat S_2(-\x)\right]^2\rangle &=&\langle \left[\hat S_3(\x) -\hat S_3(-\x)\right]^2\rangle \\
&=&\frac{2T}{D_R}\int_{D(\x)} d\xp \left[ H_1 (\xp,0) -H_2 (\xp,0) \right]
								\label{DS2-} 
\eeqa
The noise in the difference between Stokes operators
measured from symmetric pixels in general does not vanish, due to the lack of symmetry $\x, \om \to -\x ,-\om$
in the gain functions. In turns, this reflects the effect of spatial walk-off between the ordinary/extraordinary
beams (described by the term proportional to $q_y$ in the phase mismatch function \ref{delta}) and
the group velocity mismatch beteween the two waves (described by the term proportional to $\om$ in \ref{delta}).
\subsubsection{Narrow-band frequency filtering results}\label{secnarrow}

\begin{figure}[h]
\centerline{\scalebox{.50}{\includegraphics*{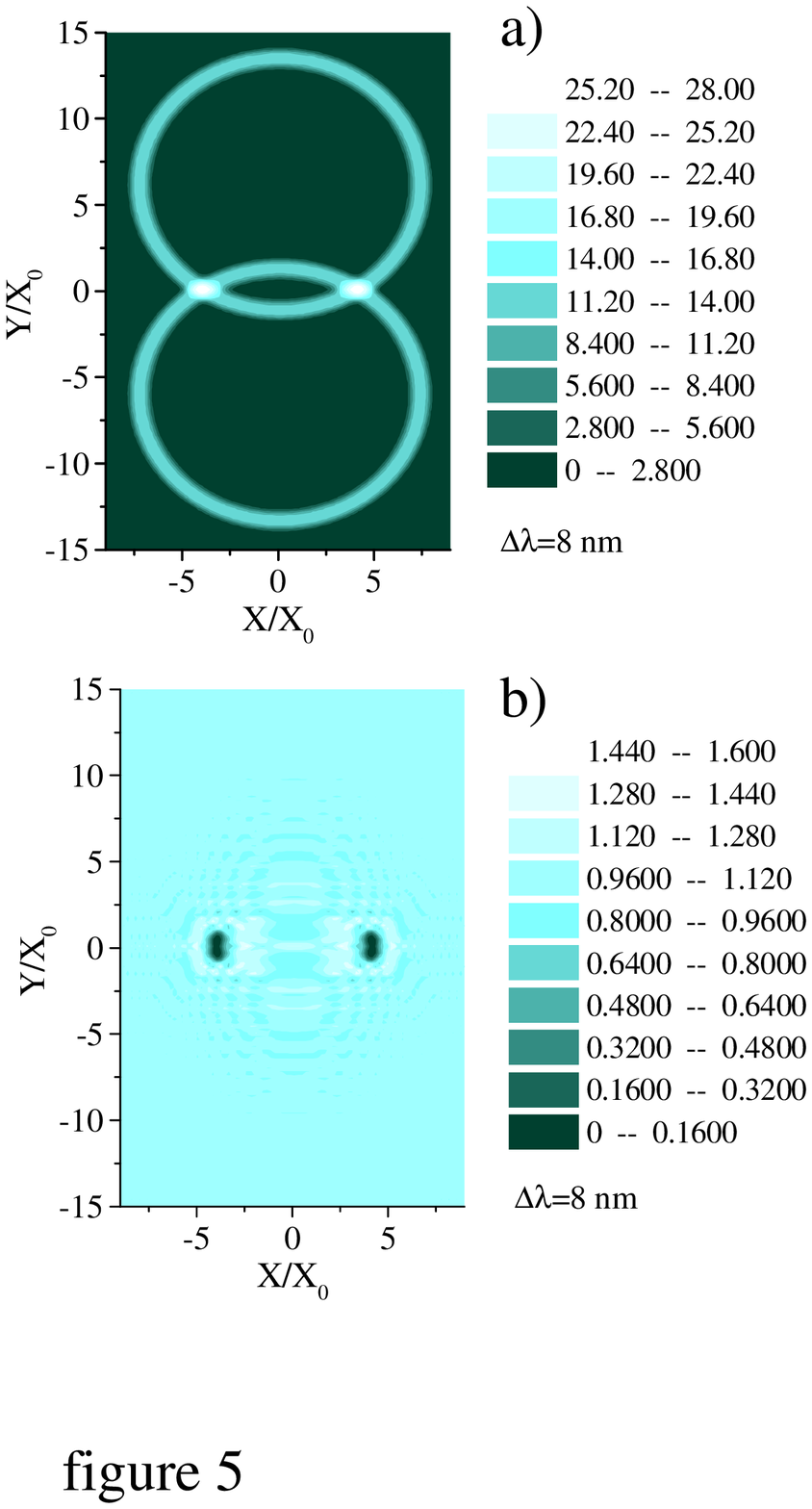}}}
\caption{a) Far field photon number distribution of the down-converted field. b)Distribution of the noise
in the difference between $\hat{S}_2$ measured from symmetric portions 
of the beam cross-section, scaled to the shot-noise level. The distribution for $\hat{S}_3$ is identical. 
A step function frequency filter $\Delta \lambda =8$ nm wide, centered around the degenerate frequency 
is used in both plots. $\sigma=2$}
\label{fig5}
\end{figure}
Part (b) of Fig.\ref{fig5} shows
a typical result for the noise in the difference between Stokes operators measured from small (but larger
than
$D_R$) symmetric
portions of the far-field. Precisely, the figure shows
$\langle \left[\hat{S}_2 (\x)-S_2(-\x)\right]^2 \rangle=
\langle \left[\hat{S}_3 (\x)-S_3(-\x)\right]^2\rangle$, scaled to the shot noise level, represented 
by $\langle \left[\hat{S}_0 (\x) + S_0(-\x)\right] \rangle$, as a function of the transverse
coordinate $\x=(x,y)$ scaled to $X_0$. Parameters in this plot are those 
of a 2 mm long BBO  crystal, cut at $49.6$ degrees for degenerate 
type II phase matching at $702\rm{ nm}$.  For comparison, part (a) of the figure shows the mean photon number
distribution in the  far field. The numbers associated with the  scale in (a) represent  the
number of photons detected over a resolution area $D_R$ and over a crystal coherence time, that is the mean 
{\em photon number per mode}.
Both plots have been obtained by filtering the emitted frequencies
in a bandwidth $\Delta \lambda= 8$ nm wide around the degenerate frequency, by means of a step function filter.
Precisely, we let 
\be
 \hat{A}_i (\x, \om) \to \sqrt{f(\om)} \hat{A}_i (\x, \om) +\sqrt{1-f(\om)} \hat v_i (\x, \om) \quad i=o,e\; ,
\label{filter}
\ee
where $\hat v_i (\x,\om)$ are vacuum field operators uncorrelated from the signal and idler fields $\hat{A}_i (\x,\om)$,
and the filter function $f(\om)$ is this case the step function $f(\om)=1$ for 
$\om \epsilon [-\frac{\Delta \om}{2},\frac{\Delta \om}{2}]$, $f(\om)=0$ elsewhere.

\begin{figure}[h]
\centerline{\scalebox{.50}{\includegraphics*{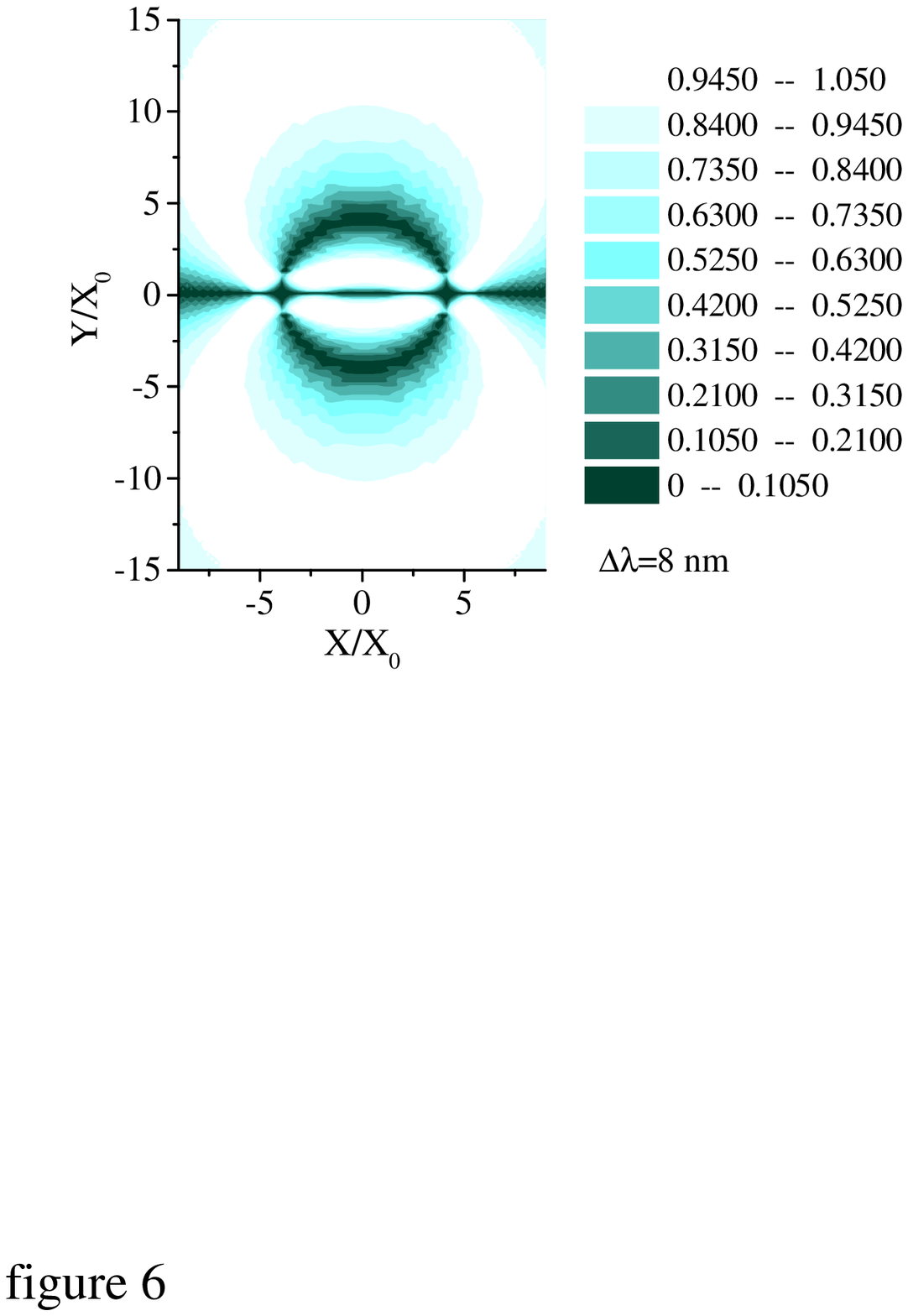}}}
\caption{Degree of polarization of the  light downconverted by a BBO crystal.
Far field distribution of $\langle \hat S_1 (\x) \rangle /\langle \hat S_0 (\x) \rangle $. 
Same parameters as in Fig.\ref{fig5} }
\label{fig6}
\end{figure}

\begin{figure}[h]
\centerline{\scalebox{.50}{\includegraphics*{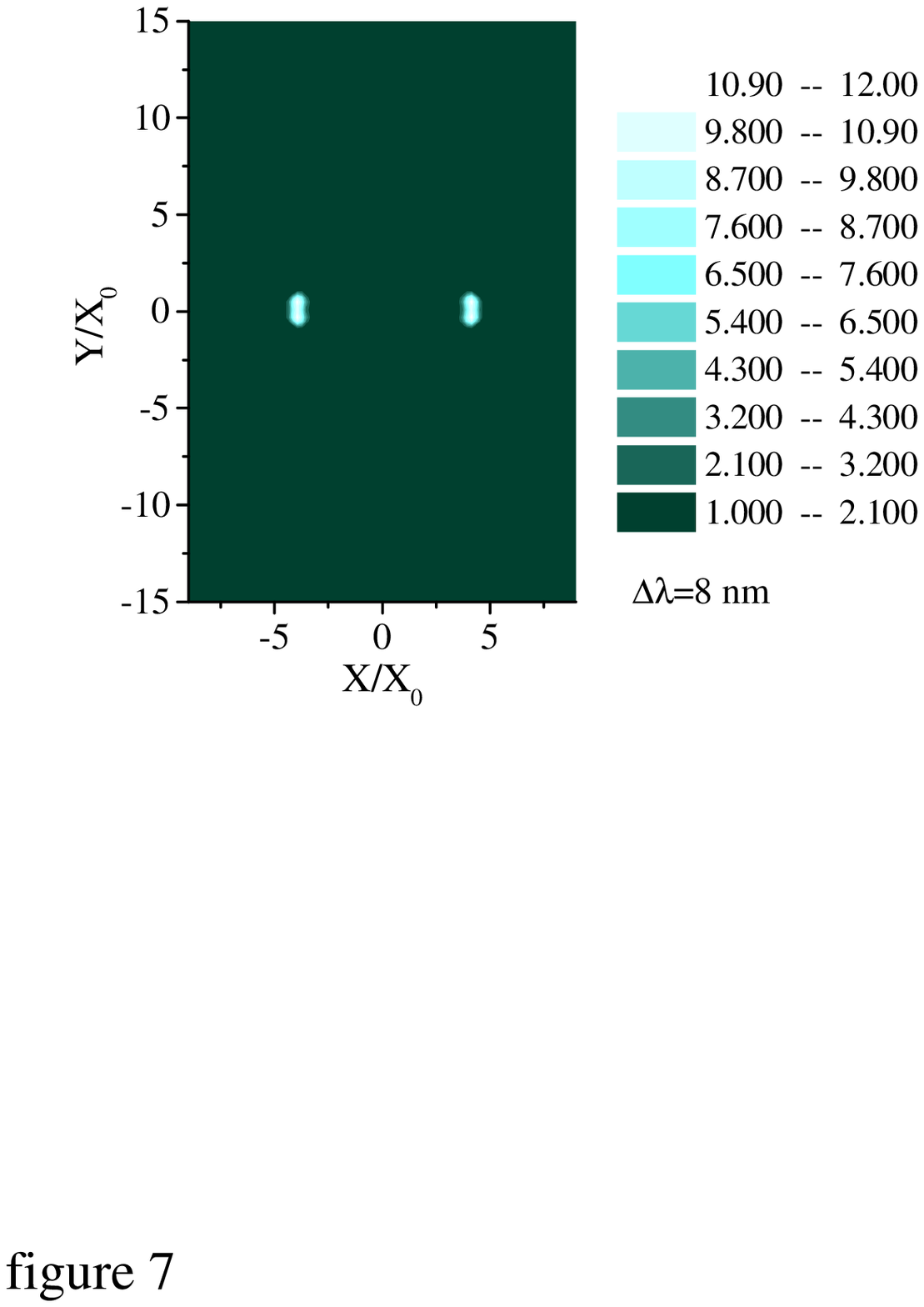}}}
\caption{Noise in the measurement of Stokes parameters of  the downconverted light by a BBO crystal.
Far field distribution of $ \langle [\delta \hat S_2(\x) ] ^2 \rangle /\langle \hat S_0 (\x) \rangle=
 \langle \hat [\delta S_3(\x)]^2 \rangle /\langle \hat S_0 (\x) \rangle$. Same parameters as in Fig.\ref{fig5} }
\label{fig7}
\end{figure}

\begin{figure}[h]
\centerline{\scalebox{.60}{\includegraphics*{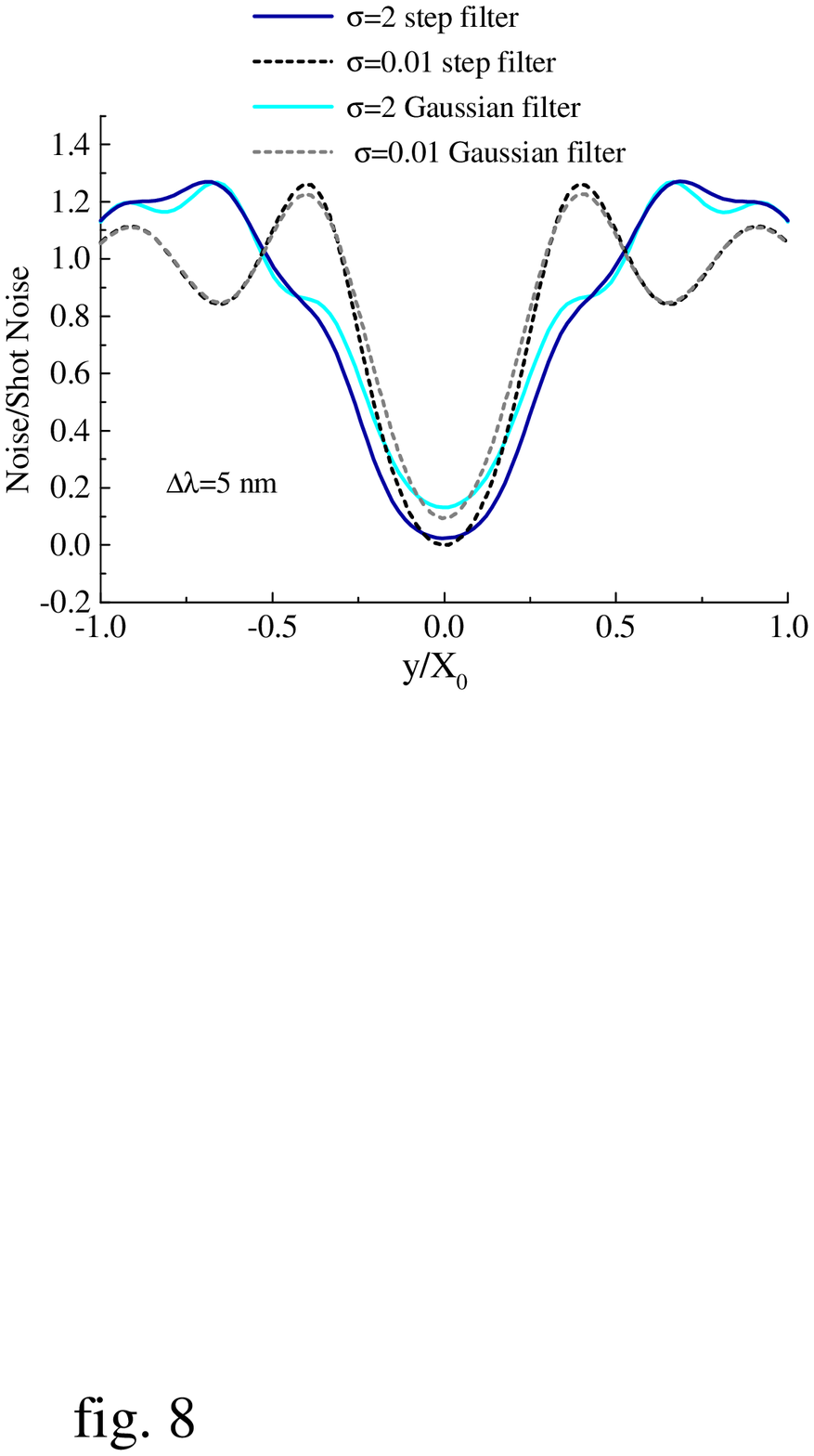}}}
\caption{Noise
in the difference between $\hat{S}_2$ ($\hat{S}_3$)  measured from symmetric portions 
of the beam cross-section, scaled to the shot-noise level, as a function of the vertical coordinate $y$ along
the maximum gain circle for the degenerate frequency. Dashed lines: $\sigma=0.01$, 
solid lines $\sigma=2$
Light lines: Gaussian frequency filter, of FWHM = 5nm. Dark lines:
step function filter 5nm wide.}
\label{fig8}
\end{figure}

In plot (b) we see clearly two large dark zones, in correspondence of the intersections of 
the  emission cones, where the Stokes operator correlation is almost perfect. 
Out of these regions, basically no spatial correlation at the quantum level exists
for Stokes operator $S_2$ and $S_3$. \\
Remarkably, at the intersection of the two degenerate emission cones, the light is completly unpolarized. Figure \ref{fig6}
shows the distribution of $\langle \hat S_1 (\x) \rangle /\langle \hat S_0 (\x) \rangle $ 
in the transverse far-field plane, showing that it vanishes at the emission ring intersection; recalling 
that the mean value of $\hat S_2$ and $\hat S_3$ is zero everywhere, this means a vanishing degree of polarization
in these regions. Moreover, in these regions a measurement of Stokes parameters
over a single detection pixel is very noisy, as shown by figure~\ref{fig7}, which plots the distribution
of $ \langle [\delta \hat S_2 (\x)]^2 \rangle =\langle [\delta \hat S_3 (\x)]^2 \rangle $, scaled to the shot noise level
$\langle \hat S_0 (\x) \rangle $. In this plot the uniform dark background correponds to the shot noise level,
while the bright spots correspond to a noise level 10-12 times larger than the shot noise.\\
Similar results are obtained in any gain regime. In the small gain limit, the noise statistics 
associated to a measurement over a single pixel becomes essentially 
Poissonian, but the correlation between Stokes parameters measured from symmetric 
pixels is basically the same as in the high gain regime. Fig. \ref{fig8} compares the 
noise in the difference between Stokes parameters
measured from symmetric pixels in the small and high gain regimes, plotted as a function of the vertical coordinate along 
the circle of maximum gain for the degenerate frequency. The dashed lines were obtained with $\sigma=0.01$,
corresponding to a mean photon number per mode $\approx 10^{-4}$,  
the solid lines with $\sigma=2$,
corresponding to a mean photon number per mode $\approx 15$ (see Fig.\ref{fig5}a),. 
 In this plot, the two dark lines
are, as usual, obtained by filtering the frequencies with a step function ($\Delta \lambda =5$nm). For comparison, 
 the two light lines show the results obtained 
by means of a more realistic frequency filter, with a Gaussian profile. Precisely,
we take the filter function in Eq.(\ref{filter}) as 
$f(\om)= \exp{\left[-(\om^2 4{\rm ln}2 )/(\Delta \om^2) \right] }$ where $\Delta \om$ is the full width at half maximum 
(FWHM),
and corresponds to an interval in wavelengths of 5nm.
In this
case, losses introduced by the Gaussian shape of the filter slightly deteriorates the correlation.

The results described above were obtained
by exploiting a trick commonly used in the experiments performed in the single photon pair regime
( for example in the experiment of \cite{kwiat}), in order to partially compensate
for the temporal and spatial walk-off of the down-converted beams. In the regime of single photon pair
production, the ordinary and extraordinary photons can be in principle distinguished because of their different
group velocities inside the crystal, and because of their offsets in propagation directions due to walk-off
effects inside the crystal. The mere existence of this possibility is detrimental for the entanglement of the state. 
As it is discussed in details in Appendix A, in the general case (arbitrary number of down-converted
photons), the group velocity mismatch and the spatial walk-off are responsible for the 
appearance of a propagation phase factor that lower the value of the correlation function between Stokes 
operators measured
from symmetric regions. In principle, this problem can be solved by using a vey narrow frequency filter, and
by performing the measurement over  narrow regions centered around the ring intersections. However, this lower 
the efficiency of the set up. An other possibility is to insert 
 a second 
crystal, after the pump beam has been removed, and after the field polarization has been   
rotated by $90^0$ (see Fig.\ref{figsetup}).   In this way, the slow and fast wave in the first crystal become  the fast and slow
wave, respectively,  in the second crystal, and the direction of walk-off is reversed. At difference from the single photon pair regime, the correlation is optimized 
when the  length of this 
second crystal is chosen as 
\be
l^\prime_c= l_c \frac{\tanh \sigma } {2\sigma} \: ,
\label{elleprimo}
\ee
where $\sigma$ is the linear gain parameter, 
proportional to the 
pump amplitude and to the first crystal length (see Appendix A). 
The fact that the optimal length of the compensation crystal decreases with increasing gain
can be understood as following \cite{John,Bahaa}: in the regime of single photon pair production 
(limit $\sigma \to 0$), the photon
pair can be produced at any point along the crystal length with uniform probability, so that the average 
temporal delay of the two photons due to the group velocity mismatch are those corresponding
to half of the crystal length, and best compensation is achieved for $l^\prime_c=\frac{l_c}{2} $. In the large
gain regime, more and more photon pairs are produced towards the end of the crystal (the number
of down-converted photons increases exponentially with the crystal length), so that walk-off effects are best
compensated by a shorter crystal, whose length is given by formula (\ref{elleprimo}).\\
\begin{figure}[h]
\centerline{\scalebox{.60}{\includegraphics*{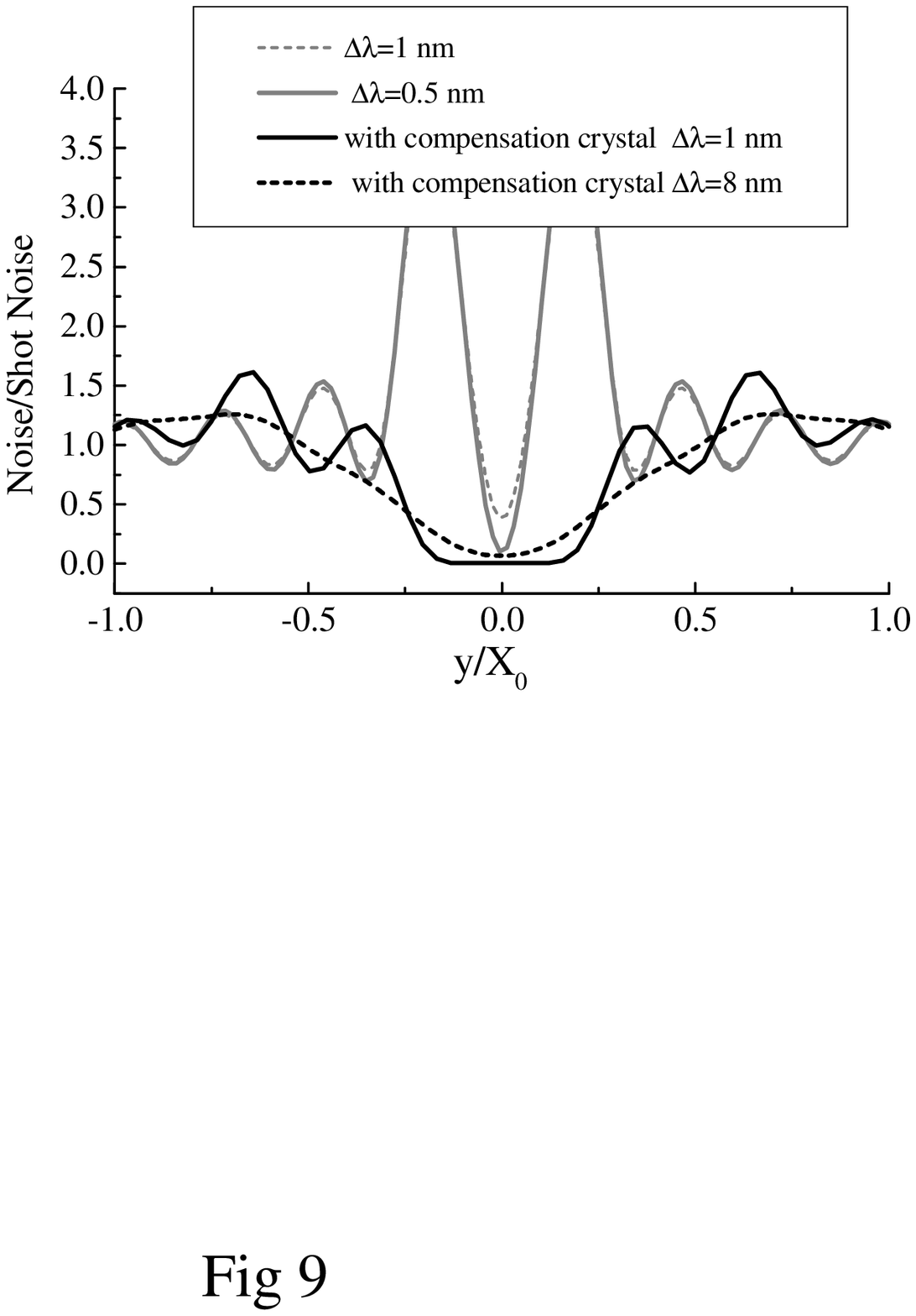}}}
\caption{Effect of the compensation crystal. Noise
in the difference between $\hat{S}_2$   measured from symmetric pixels, scaled to the shot-noise level.
$y$ is the far-field vertical position along
the maximum gain circle. Gray lines: without compensation crystal, dashed gray line $\Delta \lambda=1$nm, solid gray line
$\Delta \lambda=0.5$nm. Black lines: with optimal compensation crystal, solid black line $\Delta \lambda=1$nm,
dashed black line $\Delta \lambda=8$nm.}
\label{fig9}
\end{figure}
When this kind of optimization is not possible, our calculations show 
that similar results can be obtained
by a narrow-band temporal and spatial filtering, and/or by using crystals that exhibit a smaller amount of  walk-off.
Figure \ref{fig9} details the role of the compensation crystal. It plots the noise in the 
difference between Stokes parameters measured from symmetric pixels as a function of the vertical coordinate
$y$ along the circle corresponding to the maximum gain at the degenerate frequency. 

\subsubsection{Broad-band frequency filtering results}\label{broad}
\begin{figure}[h]
\centerline{\scalebox{.5}{\includegraphics*{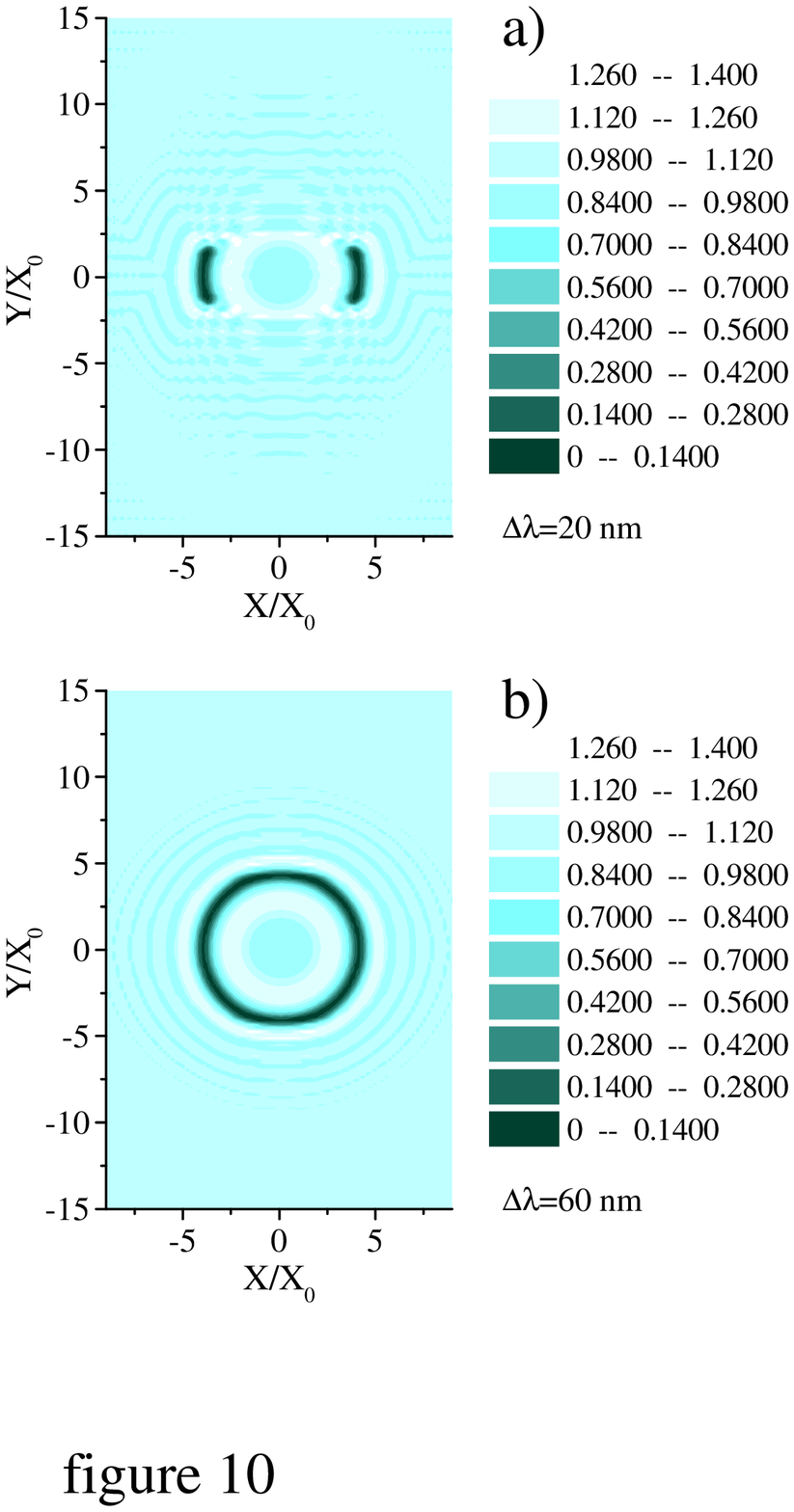}}}
\caption{Broad-band frequency filtering. Distribution of the noise
in the difference between $\hat{S}_2$ ($\hat{S}_3$)  measured from symmetric portions 
of the beam cross-section, scaled to the shot-noise level. 
In part a) a frequency filter $\Delta \lambda =20$ nm wide, centered around the degenerate frequency is used; 
in b) $\Delta \lambda=60$ nm.}
\label{fig10}
\end{figure}
\begin{figure}[h]
\centerline{\scalebox{.60}{\includegraphics*{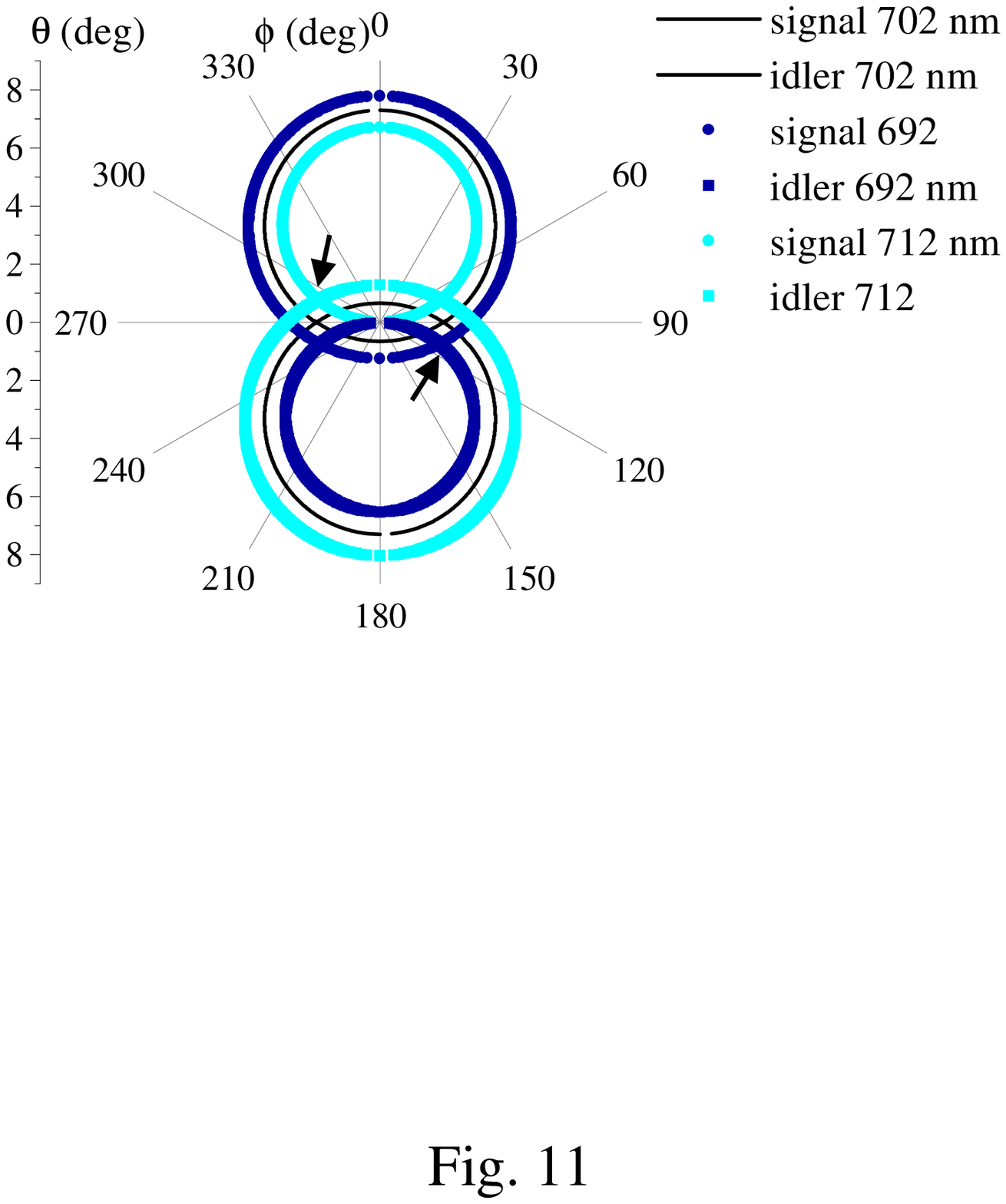}}}
\caption{Polar plot of phase matching curves in a 2 mm BBO crystal cut at 49.6 degrees. $\theta$ is the 
polar angle from the pump direction of propagation; 
$\phi$ is the azymuthal angle around the pump. Black thin curves: $ \lambda_{signal}= \lambda_{idler}= 702 nm$
Dark thick curves:   $ \lambda_{signal}= \lambda_{idler}= 692 nm $. Light thick curves:$ \lambda_{signal}= \lambda_{idler}= 712.29$ nm.
The signal is the ordinary (slow) wave, and correspond to curves in the upper half of the plot. 
Idler is the extraordinary (fast) wave.}
\label{fig11}
\end{figure}
The results described in section \ref{secnarrow}
 were obtained by using relatively narrow frequency filters (5-8 nm). Remarkably, when a broader
frequency filter is employed, the regions where Stokes parameter are correlated stretch to form a ring-shaped region
around the pump direction (see Fig. \ref{fig10}). 
This kind of shape can be understood by considering the geometry
of the downconversion cones emitted at the various frequencies by a BBO crystal. Figure \ref{fig11} is a polar plot of the
phase matching curves (geometrical loci of the phase matched modes), with $\theta$ being the polar angle from the
pump direction of propagation ($z$ axis) and $\phi$ the azymuthal angle around $z$. 
In this plot the same color identifies
the same emission wavelength; dark/light thick curves correspond to two conjugate wavelengths, 
while the thin black curves are the
two emission cones at the degenerate wavelength.
The signal (ordinary) wave emission curves are those in the upper half of the plot. 
When considering the intersection points of a dark circle with a light circle, 
which correspond to two conjugate wavelenghts,
we cannot expect any kind of entanglement, since photons arriving in these positions are clearly distinguishable
by their different frequencies. Let us consider, instead, one of the intersection of the light curves (e.g the one pointed
by the arrow in the plot). Here ordinary and extraordinary photons arrive with identical probability,
and have the same wavelength. As a consequence, the photon polarization is undetermined,
and the light is completely unpolarized. However, each time an ordinary (extraordinary) 
photon arrives at this position, an extraordinary (ordinary) 
photon, at the conjugate wavelength,  will be found at the symmetric position. This
corresponds to the intersection of the two dark curves, indicated
in the plot by the second arrow. Hence, when considering photodetection from the two regions indicated by 
the arrows in the plot we can expect a 
high degree of polarization entanglement. The same reasoning can be made for any of 
intesection of circles corresponding
to the same wavelength (light with light, dark with dark and thin black with thin black). 
By connecting all these points toghether,
we can for example recognize the geometrical shape of the dark regions in Fig.\ref{fig10}a, where a high degree of correlation
in all the Stokes parameters exists. By including more frequencies, the ring-shaped region if Fig. \ref{fig10}b shows up.\\
From the mathematical point of view, the form of the region where Stokes parameters are quantum correlated
is given by  the solution of the equation:
\be
\Delta(\q,\om) =\Delta(-\q, -\om)=0 \label{ellipse}
\ee
with $\x= 2\pi \q /(\lambda f)$. By introducing the explicit form the the phase mismatch function \ref{delta},
(\ref{ellipse}) is the equation of an ellipse (with a small eccentricity) centered around the origin, that is,
the best correlated modes form a slightly asymmetric cone around the pump direction.

\section{Polarization correlation: Quantum state formalism}
\label{Sec3}
This section is devoted to the discussion of  the problem in terms of an equivalent quantum state formalism.
Our aim is, on the one side, to give an alternative and istructive point of view on the problem, which can be compared with 
already existing quantum-state description of the problem (see e.g \cite{Bowmeester}). On the other side, we think that
this section will show how the quantum field formalism developed in the first part of this paper
 is more powerful
and strightforward, in terms of calculation efforts, than the commonly used quantum state formalism, at least
for this kind of multi-mode problems.

Equation (\ref{inout1}) defines a linear transformation acting on field operators, that maps field operators
at the entrance face of the crystal into those at the output face. 
An equivalent transformation, acting on the quantum state of the signal/idler fields
at the crystal input and mapping it into the state at the crystal output, is derived in details in Appendix B.
As described in the appendix, in order to avoid formal difficulties coming from a continuum of modes, we have
introduced  a quantisation box both in the transverse spatial domain and in the temporal domain, 
so that the continuum of spatio-temporal modes $ \q, \om$ 
is replaced by a discrete set  of modes.

When at the input of the parametric crystal
there is the vacuum state for both signal and idler fields
\be
\left| \psi \right\rangle_{in}  = \left| vac \right\rangle = \prod_{\q,\om} \left| 0; -\q,-\om \right\rangle_o   \,
\left| 0; \q,\om \right\rangle_e   \; , \label{vacuum}
\ee
we find that the output state takes the form
\beqa
\left| \psi \right\rangle_{out} &=& \prod_{\q,\om} \left\{
\sum_{n=0}^{\infty}  c_n(\q,\om) \left| n; \q,\om\right\rangle_o \,  \left| n; -\q,-\om\right\rangle_e   \right\} \;\\
c_n(\q,\om)&=& \frac{1}{\cosh r(\q,\om)} \left[\tanh r(\q,\om)\right]^n \e^{2\im n \psi (\q,\om)}
= \frac{\left[ U_o(\q,\om) V_e(-\q,-\om)\right]^n}{\left| U_o (\q,\om)\right|^{2n+1}} 
						\label{thestate1}\: ,
\eeqa
where the notation $\left| n; \q,\om\right\rangle_{o/e}$ indicates   the Fock state with n photons in mode $(\q,\om) $
of the ordinary/extraordinary polarized beam. Here the functions $U_o, V_e$ are the 
coefficients of the operator transformation (\ref{inout1}), and functions $r$ and  $\psi$  are defined
by Equations (\ref{bla2},\ref{bla3}) toghether with  (\ref{U}, \ref{V}).\\
The state (\ref{thestate1}) is clearly entangled (non factorizable) with respect to the ordinary and extraordinary
polarized beam components.

Let us focus on two conjugate modes $\q, \om$ and $-\q,-\om$ for both the ordinary and extraordinary field components.
These can be for example observed by using a narrow filter around the degenerate frequency $\om=0$ and
collecting light from two diaphragms  placed around two symmetric regions in the far field zone.
For brevity of notation, let us label these  modes with the 3D vectors $\vxi= (q_x,q_y,\om)$, 
$-\vxi= (-q_x,-q_y,-\om)$.
When restricted to these modes, the state takes the form
\beqa
\left| \psi \right\rangle_{out}^{\vxi} &=& \left\{
\sum_{n}  c_n(\vxi) \left| n; \vxi \right\rangle_o \,  \left| n; -\vxi \right\rangle_e   \right\} 
\left\{
\sum_{  n^{\, \prime }  }  c_{n^{\, \prime } } (-\vxi) 
\left| n^{\, \prime }  ; -\vxi \right\rangle_o \,  \left| n^{\, \prime }  ; \vxi \right\rangle_e   \right\} 
\label{thestate2}\\
&=& \sum_{N=0}^{\infty} \left| \phi \right\rangle_{N}^{\vxi} \label{thestateexp}\\
\left| \phi \right\rangle_{N}^{\vxi} 
&=& \sum_{m=0}^{N} \gamma_{N,m}(\vxi)   
\left| m; \vxi \right\rangle_o \,  \left| N-m; \vxi \right\rangle_e   
\left| N-m; -\vxi \right\rangle_o \,  \left| m; -\vxi \right\rangle_e  
				\label{thestateN}  
\eeqa
where the last two lines have been obtained by changing the dummy summation variables $n, n^{\, \prime }  $ into
$m=n$, $N=n+n^{\, \prime } $. \\
The state can be represented as a superposition of states with a fixed total number of photons $N$. In each N-photon
state described by Eq.(\ref{thestateN})
\beqa
\gamma_{N,m}(\vxi)   &=&
c_m(\vxi)c_{N-m}(-\vxi) \nonumber\\
&=&\frac{  [{\rm tanh}r(\vxi)]^m   [{\rm tanh}r(-\vxi)]^{N-m}  }
{    {\rm cosh} r(\vxi)\, {\rm cosh} r(-\vxi) } {\rm e}^{2\im N \psi(-\vxi)} {\rm e}^{2\im m [  \psi(\vxi)-\psi(-\vxi)  ]} 
\label{coefficients}
\eeqa
represents the probability {\em amplitude} of finding m ordinary photons, N-m extraordinary photons in mode $\vxi$,
and N-m ordinary photons, m extraordinary photons in the conjugate mode $-\vxi$. The description of the state
given by Eqs.(\ref{thestate2}-\ref{thestateN}) is a generalization of that derived in e.g.\cite{Bowmeester}. 
The main improvement
is that our description includes the effects of spatial and temporal walk-off, and allows the quantitative
evaluation of all the quantities of interest by using the parameters of a real crystal. Remarkably,
when the spatial and temporal walk-off are not taken into account, it holds
the symmetry  $(\q,\om)  \to (-\q,-\om)$. In this case, in Eq.(\ref{coefficients}) we would have
 $r(-\vxi) =r(\vxi) $ and $\psi(-\vxi)= \psi(\vxi)$, and all the coefficients $\gamma_{N,m}(\vxi)$ would
be identical for a given $N$, so that all the terms in the expansion(\ref{thestateN})  would have the same weight,
thus leading to a ``{\em maximally entangled state for polarization}"\cite{Bowmeester}. 

Coming to Stokes parameter correlation we notice the following property of the state:
\beqa
[\hat{A}^{\dagger}_o (\vxi) \hat{A}_o(\vxi) -\hat{A}^{\dagger}_e (\vxi) \hat{A}_e(\vxi)] \left| \phi \right\rangle_{N}^{\vxi} 
&=&\sum_{m=0}^{N} c_m(\vxi) c_{N-m} (-\xi) (2m-N) 
	\left| m; \vxi \right\rangle_o \,  \left| N-m; \vxi \right\rangle_e   
\left| N-m; -\vxi \right\rangle_o \,  \left| m; -\vxi \right\rangle_e  \\
&=& -[\hat{A}^{\dagger}_o (-\vxi) \hat{A}_o(-\vxi) -\hat{A}^{\dagger}_e (-\vxi) \hat{A}_e(-\vxi)] \left| \phi \right\rangle_{N}^{\vxi} 
\eeqa
By recalling the definition of the Stokes operator densities given by Eqs.(\ref{sigma1}-\ref{sigma3}) 
$\hat \sigma_1 (\vxi)= \hat{A}^{\dagger}_o (\vxi) \hat{A}_o(\vxi) -\hat{A}^{\dagger}_e (\vxi) \hat{A}_e(\vxi) $, 
$\hat \sigma_2 (\vxi)= \hat{A}^{\dagger}_o (\vxi) \hat{A}_e(\vxi) +\hat{A}^{\dagger}_e (\vxi) \hat{A}_o(\vxi) $ and
$\hat \sigma_3 (\vxi)= -i[\hat{A}^{\dagger}_o (\vxi) \hat{A}_e(\vxi) -\hat{A}^{\dagger}_e (\vxi) \hat{A}_o(\vxi)] $,
we can hence conclude  
that the state is an eigenstate of $\hat \sigma_1 (\vxi)+ \hat \sigma_1 (-\vxi)$ with
zero eigenvalue.
On the other side, we have
\be
\hat{A}^{\dagger}_o (\vxi) \hat{A}_e(\vxi) \left| \phi \right\rangle_{N}^{\vxi} 
=\sum_{m=0}^{N-1} c_m(\vxi) c_{N-m} (-\xi) \sqrt{(m+1)(N-m)} 
	\left| m+1; \vxi \right\rangle_o \,  \left| N-m-1; \vxi \right\rangle_e   
\left| N-m; -\vxi \right\rangle_o \,  \left| m; -\vxi \right\rangle_e  \; .
\ee
\beqa
\hat{A}^{\dagger}_o (-\vxi) \hat{A}_e(-\vxi) \left| \phi \right\rangle_{N}^{\vxi} 
&=&\sum_{m=1}^{N} c_m(\vxi) c_{N-m} (-\xi) \sqrt{m(N-m+1)} 
	\left| m; \vxi \right\rangle_o \,  \left| N-m; \vxi \right\rangle_e   
\left| N-m+1; -\vxi \right\rangle_o \,  \left| m-1; -\vxi \right\rangle_e  \\
&=&\sum_{l=0}^{N-1} c_{l+1}(\vxi) c_{N-l-1} (-\xi) \sqrt{(l+1)(N-l)} 
	\left| l+1; \vxi \right\rangle_o \,  \left| N-l-1; \vxi \right\rangle_e   
\left| N-l; -\vxi \right\rangle_o \,  \left| l; -\vxi \right\rangle_e  \; ,
\eeqa
where the last line has been obtained by introducing the summation index $l=m-1$.
This implies that the equation
\be
\left[\hat{A}^{\dagger}_o (\vxi) \hat{A}_e(\vxi) -\hat{A}^{\dagger}_o (-\vxi) \hat{A}_e(-\vxi)\right] 
\left| \psi \right\rangle_{out}^{\vxi} =0
\ee
is verified if and only if 
\be
c_m(\vxi) c_{N-m} (-\vxi) =c_{m+1} (\vxi) c_{N-m-1}(-\vxi) 
\label{cond1}
\ee
for all $N=0,+\infty$ and $m=0,N-1$.\\
Similar considerations for the hermitian conjugate operator 
$\hat{A}^{\dagger}_e (\vxi) \hat{A}_o(\vxi) -\hat{A}^{\dagger}_e (-\vxi) \hat{A}_o(-\vxi)$ lead to the equivalent 
condition
\be
c_m(\vxi) c_{N-m} (-\vxi) =c_{m-1} (\vxi) c_{N-m+1}(-\vxi) 
\label{cond1bis}
\ee
for all $N=0,+\infty$ and $m=1,N$.\\
Hence, the state is also an eigenstate 
of both
$\hat \sigma_2 (\vxi)- \hat \sigma_2 (-\vxi)$ and 
$\hat \sigma_3 (\vxi)- \hat \sigma_3 (-\vxi)$, with zero eingevalue, if and only if 
the conditions (\ref{cond1},\ref{cond1bis} ) are
satisfied. These conditions amount to requiring that all the coefficients in the expansion
of the N-photon state (\ref{thestateN}) are identical, and that the N photon state is a superposition
with equal probability amplitude of all the possibile partitions in m ordinary and N-m extraordinary
photons (m=0,N) in mode $\vxi$, with N-m ordinary and m extraordinary photons in the conjugate mode $-\vxi$.
This is the mathematical equivalent of the commonly used  statement ``{\em ordinary and extraordinary photons in mode $\vxi$ are not
distinguishable, but each time we have m ordinary and N-m extraordinary photon in mode $\vxi$, there
are N-m ordinary and m extraordinary photons in mode $-\vxi$}". 
For modes having a non vanishing parametric gain the conditions (\ref{cond1},\ref{cond1bis}) amount to requiring
\be
{\rm tanh} r(\vxi) e^{2\im \psi(\vxi)} ={\rm tanh} r(-\vxi) e^{2\im \psi(-\vxi)} \; , \label{cond2}
\ee
a condition that is satisfied only in the presence of the symmetry $\Delta(\q,\om)=\Delta(-\q,-\om)$. This in turns
implies the absence of spatial walk-off between the two waves(i.e. the two modes correspond to the intersection of the
down-conversion cones) and the absence of temporal walk-off (use of a narrow frequency filter and/or compensation by 
means of a second crystal).\\
Formula (\ref{cond2}) can be also written as:
\be
U(\vxi) V^*(-\vxi)= U(-\vxi) V^*(\vxi) \; .
\ee
By comparing with equation (\ref{H10-H20}), we notice that this is the condition that ensures  that the
correlation between Stokes parameter measured from symmetric pixels calculated in Section \ref{sec23}
reaches its maximum value. 
Hence, in the framework of the quantum state formalism, we start to recover  the same results of Section \ref{Stokescorr},
as it obviously must be. One could proceed further on, and derive quantitative results for the correlation, as 
those showed by Figs.\ref{fig5}-\ref{fig10}, but at this point it should be rather clear (and for sure we are
not the first ones to notice this) how the quantum state
formalism, although instructive, is cumbersome and not transparent in comparison with the quantum field formalism.

\section{Conclusions}
\label{conclusions}

In conclusion, we have shown that the polarization entanglement of photon pairs emitted in 
parametric down-conversion survives in high gain regimes, where the number of converted photons can be rather large.
In this case, it takes the form of non-classical spatial correlations of {\em all} 
light Stokes operators associated to polarization degrees of freedom. 
We have shown that in  
the regions where the two rings intersect (in a ring-shaped region around the pump direction when a 
broad frequency filter is employed) all the  Stokes  operators are highly correlated
at a quantum level, realizing in this way a macroscopic  polarization entanglement. 
Although Stokes 
parameters are extremely noisy and the state is unpolarized,
measurement of a Stokes parameter {\em 
in any polarization basis} in one far-field region determines the Stokes parameter
collected from the symmetric region, within  an uncertainty much below the standard quantum limit.\\
We call this situation ``polarization entanglement" because, on the one side, the quantum state 
derived in Section \ref{Sec3} is entangled with respect to polarization degrees of freedom, and, on the other,
because in our description there is no gap in the passage from the single photon pair regime, where 
the polarization entanglement
is a widely  accepted concept, to the multiple photon pair regime. However, we want to remark that for spontaneous
parametric down-conversion
there is no way, to our knowledge, to derive a sufficient criterium for inseparability based on the degree of correlation
of the Stokes operators, as this derived in \cite{Duan} and generalized by \cite{Bowen2}. This depends on the fact
that the average
values of commutators (and anticommutators) of Stokes operators are in this system intrinsically state dependent, 
at difference to what happens in the experiment performed in \cite{Bowen2}, where bright entangled beams were used.
Further discussion about this important point is postponed to a future publication.

We have developed a multi-mode model for spontaneous parametric down-conversion, 
both within the framework of quantum field formalism and quantum state
formalism. They are valid in any gain regime, from the single photon pair production  to the high gain regime where
the number of downconverted photons can be rather large. The model allows quantitative estimations of all
the quantities of interest, by using empirical parameters of real crystals. We hope that this description
can be a useful tool for experimentalists working in this field.

Quite interesting, and to our knowledge completely novel, are the results concerning the correlation
of Stokes parameters observed by using a broad frequency filter, described in Section \ref{broad}. They
basically show how by increasing the number of  temporal degrees of freedom in play , the number 
of spatial degrees of freedom which are simultaneously entangled increases, so that the two isolated
correlated spots in Figure \ref{fig5} become the ring shaped region of Figure \ref{fig10}, where many
symmetric spots are  correlated in pairs.

We believe that  this form of entanglement, with its increased complexity 
in terms of degrees of freedom (photon number, polarization, temporal and spatial degrees
of freedom)can be quite  promising 
for new quantum information schemes.

\section*{Appendix A} \label{appendixA}
In this Appendix we calculate the phase shift induced by the propagation 
of the down-converted fields through a compensation crystal, 
and we  evaluate the length of this second crystal necessary for optimal walk-off compensation.\\
As shown by the scheme of Fig.\ref{figsetup}, we assume that after producing down-conversion
in a first crystal (BBO1), the pump beam is eliminated. The polarisations of the downconverted beams  
is then rotated by 90 degrees,  and they pass through  
a second crystal(BBO2) of length  $l_c^\prime$, identical to the first one.\\ 
In the region between the second crystal and the lens $L$ the ordinary/extraordinary 
field operators can be written as:
\beqa
\hat{A}_{o} (\q,\om,z) &=& \hat{A}_e^{out} (\q,\om) \exp{ \left[ \im k_{oz} (\q,\om) l_c^{\prime} \right] }
\exp{ [\im \phi_{vac}(z-l_c^{\prime})] } \label{ao}\\
\hat{A}_{e} (\q,\om,z) &=& \hat{A}_o^{out} (\q,\om) \exp{\left[\im k_{ez} (\q,\om) l_c^{\prime}\right]} 
\exp{ [\im \phi_{vac}(z-l_c^{\prime})] } \label{ae} \: .
\eeqa
The first phase shift accounts for propagation inside the compensation crystal. Here
$k_{oz} (\q,\om)$, $k_{ez} (\q,\om)$  are the projections along z-axis of the ordinary/extraordinary 
wave-vectors inside the crystal, whose explicit expressions depend on the linear properties of the crystal
as described by Eq.(\ref{exp}). The second phase shift accounts for paraxial propagation in vacuum 
$\phi_{vac} (z)= (k-\frac{q^2}{2k}) z$ , $k=2\pi/\lambda$.\\
In the far field plane, all the results described in Sections \ref{sec01}, \ref{sec23} remain unchanged
provided that one makes the following substitutions:
\beqa
\U_o (\x,\om) &\to& \!\! \!\!\!\!\!\! \left. \phantom{\sum}   U_e (\q,\om) 
               \exp{\left[\im k_{oz} (\q,\om) l_c^{\prime}\right] } \right|_{q=\x \frac{2\pi}{\lambda f} }\\
\V_o (\x,\om) &\to& \!\! \!\!\!\!\!\! \left. \phantom{\sum}   V_e (\q,\om) \exp{\left[\im k_{oz} (\q,\om) l_c^{\prime}\right] } 
				\right|_{q=\x \frac{2\pi}{\lambda f} }\\
\U_e (\x,\om) &\to& \!\! \!\!\!\!\!\! \left. \phantom{\sum}   U_o (\q,\om) \exp{\left[\im k_{ez} (\q,\om) l_c^{\prime}\right]  }
			\right|_{q=\x \frac{2\pi}{\lambda f} }\\
\V_e (\x,\om) &\to& \!\! \!\!\!\!\!\! \left. \phantom{\sum}   V_o (\q,\om) \exp{\left[\im k_{ez} (\q,\om) l_c^{\prime}\right]  }
					\right|_{\q=\x \frac{2\pi}{\lambda f} } \; ,
\eeqa
where global phase factors have been omitted, since they do not affect the results. \\
This transformation leaves unchanged all the results described in Sec. \ref{sec01} (noise and correlation for
measurements of Stokes operators 0 and 1). For the second and third Stokes parameters
(Sec. \ref{sec23}) , while the transformation 
does not affect 
the amount of noise of the measurement,
given by Eqs. \ref{H1},\ref{H10}, it does affect the correlation between measurements from symmetric pixels (Eqs.
\ref{H2},\ref{H20})
\beqa
H_2(\x,0)&\to& \int \frac{d \omega}{2\pi} 2{\rm Re}\left\{  \U^*(-\x,-\omega) \U(\x,\omega ) 
\V^*(-\x,-\omega) \V (\x, \omega )  \e^{\im \phi_{c} (\x,\om)}  \right\}  \: ,\\
\phi_{c} (\x,\om) &=& \!\! \!\!\!\!\!\! \left. \phantom{\sum} \left[  k_{ez} (\q,\om) + k_{oz} (-\q,-\om) - k_{oz} (\q,\om) 
-k_{ez} (-\q,-\om) \right] l_c^{\prime}  \right|_{ \q=\x \frac{2\pi}{\lambda f}}  \\
&=& \!\! \!\!\!\!\!\! \left. \phantom{\sum}   \left[ \Delta (-\q, -\om) - \Delta (\q,\om)\right] \frac{l_c^{\prime}  }{l_c}
\right|_{ \q=\x \frac{2\pi}{\lambda f}} \: .
									\label{phicomp}
\eeqa
On the other side, by using the explicit expression of the gain functions in Eqs. (\ref{U},\ref{V}), we have
\be
{\rm arg}\left\{ \U^*(-\x,-\omega) \U(\x,\omega ) 
\V^*(-\x,-\omega) \V (\x, \omega )  \right\} =\!\! \!\!\!\!\!\! 
\left. \phantom{\sum }2\psi(\q,\om) -2\psi(-q,-\om) \right|_{ \q=\x \frac{2\pi}{\lambda f}} \, 
\label{deltapsi}
\ee
with
\beqa
2\psi(\q,\om)&=& 
\tan^{-1}\left\{ \Delta(\q,\om) \frac{{\rm tanh} \Gamma (\q,\om) }{2\Gamma(\q,\om)}     \right\}  \\
&\approx& \Delta(\q,\om) \frac{ {\rm tanh} \sigma}{2\sigma} \: .
\eeqa
The last line has been obtained by taking the limit $\Delta (\q,\om) \ll 1$;  this  is meaningful
since the most important contribution to the correlation function is given by phase matched modes.
The phase factor (\ref{deltapsi}) can be partially compensated
by the phase shift induced by propagation in the second crystal (\ref{phicomp}). Best compensation is achieved
for
\be
\frac{l_c^{\prime}  }{l_c}= \frac{ {\rm tanh} \sigma}{2\sigma} 
			\label{elleprimo2}
\ee 
In this conditions the value of the correlation between measurements from symmetric pixel (the value
of the function $H_2$) is maximized by the presence of a compensation crystal.
\section*{Appendix B}
Equation (\ref{inout1}) defines a linear transformation acting on field operators, that maps field operators
at the entrance face of the crystal into those at the output face. The aim of 
this appendix is to find  is an equivalent transformation acting on the quantum state of the signal/idler fields
at the crystal input and mapping it into the state at the crystal output.\\
In order to avoid formal difficulties coming from a continuum of modes, we introduce a quantisation box
of side $b$ in the transverse plane, with periodic boundary conditions. In this way the continuum of wave-vectors
$\q$ is replaced by a set of discrete wave vectors $q_{\vec{l}}= (l_x \vec{u}_x + l_y \vec{u}_y)\frac{2\pi }{b} $
$l_x,l_y=0, \pm 1, \pm 2 ...$. In the same way, we introduce a quantization box in the time domain of length
$T$, with periodic boundaries, so that we need to consider only 
 a discrete set of temporal frequencies $\om_p =   p \frac{2\pi}{T}$ $p=0, \pm 1 ..$.
The free field commutation relation (\ref{commutator}) are thus replaced by their discrete version
$\left[   \hat{A}_i (\q_{\vec{l}}, \om_p), \hat{A}_j^\dagger (\q_{\vec{m}}, \om_s)\right] = \delta_{i,j}
\delta_{l_x,m_x}\delta_{l_y,m_y}\delta_{n,s}\quad$, $i,j=0,e$.\\
For brevity of notation, in the following we shall indicate the spatio-temporal mode $\q_{\vec{l}}, \om_p$
with the three dimensional vector $\vxi$, and we shall not write explicitly the modal indices.

The input/output transformation (\ref{inout1}) can be written in a equivalent way as:
\be
\hat{A}_i (\vxi)=\hat R^{\dagger} \hat{A}_i^{in} (\vxi) \hat R, 
			\label{inoutop}
\ee
with
\be
\hat R=\hat R_0 \hat R_1 \hat R_2 \; ,
\ee
and
\beqa
\hat R_0 &=& \exp\left\{ \im \sum_{\vxi} \left[ \psi (\vxi)+\varphi (\vxi)\right] \hat{A}^\dagger_o (\vxi)\hat{A}_o (\vxi)+
\left[ \psi (\vxi)-\varphi (\vxi)\right] \hat{A}^\dagger_e (-\vxi)\hat{A}_e (-\vxi)\right\} 
				\label{R0}\\
\hat R_1 &=& \exp\left\{ \sum_{\vxi}  r (\vxi)\left[ \hat{A}^\dagger_o (\vxi)\hat{A}^\dagger_e (-\vxi)- 
				\hat{A}_o (\vxi)\hat{A}_e (-\vxi)\right] \right\} \: ,
				\label{R1}\\
\hat R_2 &=& \exp\left\{ \im \sum_{\vxi}  \theta (\vxi) 
\left[ \hat{A}^\dagger_o (\vxi)\hat{A}_o (\vxi)+ \hat{A}^\dagger_e (-\vxi)\hat{A}_e (-\vxi)\right]  \right\} 
				\label{R2}           
\eeqa
with functions $\psi(\vxi), \varphi(\vxi), r(\vxi), \theta(\vxi)$ defined  by Eqs. (\ref{bla2},\ref{bla3}),
toghether with (\ref{U},\ref{V}, \ref{phi}). \\
In order to demonstrate the ansatz (\ref{inoutop}), we first notice that the action of operators 
$\hat R_0$ and $\hat R_2$ on field
operators  corresponds to  phase rotations. For any operator $c$, for which $[c,c^\dagger]=1$, we have
\be
\e^{-\im s  c^\dagger c}    c \e^{\im s  c^\dagger c}   =\e^{\im s} c  \; .
\ee
As a consequence, 
\beqa
\hat R_0^\dagger \hat{A}_o (\vxi) \hat R_0 &=& \hat{A}_o (\vxi)\e^{ \im \left[ \psi (\vxi)+\varphi (\vxi)\right] } \\
\hat R_0^\dagger \hat{A}_e (\vxi)\hat R_0 &=& \hat{A}_e (\vxi)\e^{ \im \left[ \psi (-\vxi)-\varphi (-\vxi)\right] } 
\: .
\eeqa
Operator $\hat R_1$ is the product of an infinity of two mode squeezing operators, each of them acting on the couple
of modes $(\vxi)$ in the signal beam and $(-\vxi)$ in the idler beam. For any couple of independent boson operators
$c_1$, $c_2$, and for $r$ real,  it holds the identity
\be
e^{-r \left[   c_1^\dagger c_2^\dagger - c_1c_2 \right] } c_1 e^{+r \left[   c_1^\dagger c_2^\dagger - c_1c_2 \right] } 
=c_1 \cosh{r} + c_2^\dagger \sinh{r}\; .
\ee
Hence, letting $c_1 \to \hat{A}_o (\vxi) $, $c_2 \to \hat{A}_e (-\vxi) $, we have 
\beqa
\hat R_1^\dagger \hat{A}_o (\vxi)\hat R_1 &=& \hat{A}_o (\vxi)\cosh{r(\vxi )} + \hat{A}_e^\dagger (-\vxi)\sinh{r(\vxi)}\\
\hat R_1^\dagger \hat{A}_e (\vxi)\hat R_1 &=& \hat{A}_e (\vxi)\cosh{r(-\vxi)} + \hat{A}_o^\dagger (-\vxi)\sinh{r(-\vxi)}\; .
\eeqa
Finally, by letting also the operator $\hat R_2$ act:
\beqa
\hat R_2^\dagger \hat R_1^\dagger \hat R_0^\dagger \hat{A}_o (\vxi)\hat R_0 \hat R_1 \hat R_2
&=& \e^{ \im \left[ \psi (\vxi)+\varphi (\vxi)\right] } \left\{
\hat{A}_o (\vxi)\cosh{r(\vxi)} \e^{ \im \theta (\vxi)} 
+ \hat{A}_e^\dagger (-\vxi)\sinh{r(\vxi)} \e^{ -\im \theta (\vxi)} \right\}
								\\
&=&\e^{ \im \varphi (\vxi)} \left\{
\hat{A}_o (\vxi)U(\vxi)
+ \hat{A}_e^\dagger (-\vxi)V(\vxi)\right\} 
\eeqa
where in passing from the first to the second line we used the relation (\ref{bla3}), 
which is a consequence of the unitarity
of the transformation (\ref{inout1}). Moreover, we have
\beqa
\hat R_2^\dagger \hat R_1^\dagger \hat R_0^\dagger \hat{A}_e (\vxi)\hat R_0 \hat R_1 \hat R_2
&=& \e^{ \im \left[ \psi (-\vxi)-\varphi (-\vxi)\right] } \left\{
\hat{A}_e (\vxi)\cosh{r(-\vxi)} \e^{ \im \theta (-\vxi)} \right. \nonumber \\
&+& \left.  \hat{A}_o^\dagger (-\vxi)\sinh{r(-\vxi)} \e^{ -\im \theta (-\vxi)} \right\} \\
&=&\e^{ -\im \varphi (-\vxi)} \left\{
\hat{A}_e (\vxi)U(-\vxi)
+ \hat{A}_o^\dagger (-\vxi)V(-\vxi)\right\} \: .
\eeqa
Finally, taking into account the relation (\ref{bla2}), which is again a consequence of the unitarity
of the transformation (\ref{inout1}), we recover the input/output transformation (\ref{inout1}).

Any quantum mechanical expectation value of the output operators (mean values, correlation functions etc.)
taken on the input state,
is equivalent to the quantum mechanical expectation value of the input operators taken on
the transformed state:
\be
\left| \psi \right\rangle_{out} = \hat R \, \left| \psi \right\rangle_{in}  
\ee
In the following we shall derive the form of the output state, when at the input of the parametric crystal
there is the vacuum state for both signal and idler fields.
\be
\left| \psi \right\rangle_{in}  = \left| vac \right\rangle = \prod_{\vxi} \left| 0; \vxi \right\rangle_o   \,
\left| 0; -\vxi \right\rangle_e   \; ,
\ee
where the notation $\left| n; \vxi \right\rangle_{o/e}$ indicates   the Fock state with n photons in mode $(\vxi)$
of the ordinary/extraordinary polarized beam.\\
First of all we notice that the operator $\hat R_2$ has no effect on the vacuum state, corresponding to a 
phase rotation of the vacuum. For what concern operator $\hat R_1$, by using proper operator ordering techniques
(see e.g. \cite{Barnett-Radmore}) pag. 75), it can be recasted in the following form ({\em disentangling theorem})
\beqa
\hat R_1 &=& \prod_{\vxi} \left\{ \e^{G(\vxi)\hat{A}_o^\dagger (\vxi)\hat{A}_e^\dagger (-\vxi)}
\e^{-g(\vxi)\left[\hat{A}_o^\dagger (\vxi )\hat{A}_o(\vxi) +  \hat{A}_e^\dagger (-\vxi)\hat{A}_e (-\vxi)  +1\right] } 
\e^{-G(\vxi)\hat{A}_o(\vxi)\hat{A}_e (-\vxi)} \right\}    
			\\
G(\vxi)&=& \tanh [r(\vxi)]  \\
g(\vxi)&=& \log \{ \cosh [r(\vxi)]  \}
\eeqa
By letting this operator acting on the vacuum state
\be
\hat R_1 \left| vac \right\rangle = \prod_{\vxi} \frac{1}{\cosh[r(\vxi)] }  
\sum_{n=0}^{\infty}  \left[\tanh{r(\vxi)} \right]^n
\left| n; \vxi \right\rangle_o   \left| n; -\vxi \right\rangle_e   \; ,
\ee
where the usual expansion of the exponential operator, 
$exp{\hat M}= \sum_{n=0}^{\infty} \frac{\hat M^n}{n!}$, has been used , toghether with the standard action of boson
creation operators on Fock states. 
Finally, by adding the action of operator $\hat  R_0$,
\be 
\hat R_0 \hat R_1 \left| vac \right\rangle = \prod_{\vxi} \frac{1}{\cosh[r(\vxi)]}   
\sum_{n}  \left[\tanh{r(\vxi)} \right]^n \e^{2\im n \psi (\vxi)}
\left| n; \vxi \right\rangle_o   \left| n; -\vxi \right\rangle_e   \; ,
\ee
the output state can be written in the form
\beqa
\left| \psi \right\rangle_{out} &=& \prod_{\vxi} \left\{
\sum_{n}  c_n(\vxi)\left| n; \vxi \right\rangle_o \,  \left| n; -\vxi \right\rangle_e   \right\} \;\\
c_n(\vxi)&=& \frac{1}{\cosh r(\vxi)} \left[\tanh r(\vxi)\right]^n \e^{2\im n \psi (\vxi)}
= \frac{\left[ U_o(\vxi)V_e(-\vxi)\right]^n}{\left| U_o (\vxi)\right|^{2n+1}}
\eeqa

\acknowledgements{This work was carried out in the framework of the EU project QUANTIM (Quantum Imaging).
Two of us (RZ and MSM) 
acknowledge financial support from the Spanish MCyT project BFM2000-1108}

%

\end{document}